# From climate models to planetary habitability: temperature constraints for complex life

**Laura Silva[1], Giovanni Vladilo[1], Patricia M. Schulte[2], Giuseppe Murante[1], Antonello Provenzale[3]**

**Abstract:** In an effort to derive temperature based criteria of habitability for multicellular life, we investigated the thermal limits of terrestrial poikilotherms, i.e. organisms whose body temperature and the functioning of all vital processes is directly affected by the ambient temperature. Multicellular poikilotherms are the most common and evolutionarily ancient form of complex life on earth. The thermal limits for the active metabolism and reproduction of multicellular poikilotherms on earth are approximately bracketed by the temperature interval $0^{o}$C $\leq T \leq 50^{o}$C. The same interval applies to the photosynthetic production of oxygen, an essential ingredient of complex life, and for the generation of atmospheric biosignatures observable in exoplanets. Analysis of the main mechanisms responsible for the thermal thresholds of terrestrial life suggests that the same mechanisms would apply to other forms of chemical life. We therefore propose a habitability index for complex life, $h_{050}$, representing the mean orbital fraction of planetary surface that satisfies the temperature limits $0^{o}$C $\leq T \leq 50^{o}$C. With the aid of a climate model tailored for the calculation of the surface temperature of Earth-like planets (ESTM), we calculated $h_{050}$ as a function of planet insolation, $S$, and atmospheric columnar mass, $N_{atm}$, for a few earth-like atmospheric compositions with trace levels of $CO_2$. By displaying $h_{050}$ as a function of $S$ and $N_{atm}$, we built up an atmospheric mass habitable zone (AMHZ) for complex life. At variance with the classic habitable zone, the inner edge of the complex life habitable zone is not affected by the uncertainties inherent to the calculation of the runaway greenhouse limit. The complex life habitable zone is significantly narrower than the habitable zone of dry planets. Our calculations illustrate how changes in ambient conditions dependent on $S$ and $N_{atm}$, such as temperature excursions and surface dose of secondary particles of cosmic rays, may influence the type of life potentially present at different epochs of planetary evolution inside the AMHZ.

[1] National Institute for Astrophysics, INAF-OATs, Trieste, Italy
[2] Department of Zoology, University of British Columbia, Vancouver (BC), Canada
[3] Institute of Geosciences and Earth Resources, CNR, Pisa, Italy

## Introduction

The search for life in the Universe is one of the main goals of astrobiology (Des Marais et al. 2003, Chyba and Hand 2005). A possible observational strategy for the quest for life outside the Solar System consists of searching for biosignatures in the atmospheres of extrasolar planets (Kaltenegger et al. 2002, Des Marais et al. 2002). Based on example of the Earth, the most promising targets for this search are terrestrial type exoplanets, i.e. rocky planets with sizes and mass comparable to those of the Earth. Terrestrial type exoplanets seem to be quite numerous, once the present-day statistics is corrected for the observational bias that affect the detection of small-size planets (Mayor et al. 2011, Foreman-Mackey et al. 2014). However, the measurement of atmospheric biosignatures in small size exoplanets is still beyond observational capabilities (Hedelt et al. 2013) and the interpretation of atmospheric features in terms of chemical disequilibrium driven by biological activity is uncertain (Segura et al. 2007, Wordsworth and Pierrehumbert 2014, Harman et al. 2015, Narita et al. 2015). In addition to these difficulties, a critical aspect of the search for life outside Earth is the lack of a consolidated theory on the origin of terrestrial life (e.g. Delaye and Lazcano 2005). As a result, it is hard to make predictions about the potential presence of life outside Earth even in Solar System planets/satellites and, a fortiori, in extrasolar planetary systems. Future breakthroughs in our observational capabilities and in our understanding of life's origins may change this current state of affairs. In the meantime, studies of exoplanets should focus on the study of the planet's ambient conditions, rather than searching for actual signatures of life. By exploring the ambient conditions we can cast light on a planet's habitability, i.e. its capability of hosting life. The habitability can be translated into a set of planetary conditions that can be tested more easily than the actual presence of life. Searches for exoplanets that satisfy such conditions can be performed even with the modest observational data available for terrestrial type targets. Moreover, our lack of understanding of life's origins does not prevent investigation of the conditions of habitability, which are independent of the actual emergence of life in a given planet. Therefore, the search for habitable exoplanets is a necessary step to select optimal targets for future, time-consuming searches of biosignatures.

The most commonly adopted requirement of planetary habitability is the liquid water criterion. This criterion has been adopted since the early studies of planetary habitability, well before the discovery of extrasolar planets (e.g. Rasool and DeBergh 1970, Hart 1979). Besides the obvious importance of water in terrestrial life, the adoption of the liquid water criterion is motivated by the requirement of a liquid solvent in biochemical processes, together with the special properties of water and the high cosmic abundance of its atomic constituents. Assuming that water is present on the planet, the liquid water criterion constrains the physical state of the environment. In spite of its conceptual simplicity, the practical implementation of this criterion requires complex calculations performed with planetary climate models (Provenzale 2014). To simplify the calculations, the limits of planetary habitability are often studied with 1D atmospheric models (Kasting 1988, Kasting et al. 1993; Kasting and Catling 2003, Pierrehumbert and Gaidos 2011, von Paris et al. 2013, Kopparapu et al. 2013), and more rarely with 3D models (e.g. Williams and Pollard 2002, Leconte et al. 2013, Yang et al. 2014). Another approach to simplify the climate calculations consists of adopting energy balance models (EBMs) with an analytical treatment of the energy transport (North et al. 1981; Pierrehumbert 2010). Thanks to their low computational cost, EBMs have been used to explore the habitability of terrestrial planets with different axis obliquities, eccentricities, rotation rates and stellar insolation (Williams and Kasting 1997; Spiegel et al. 2008, 2009, 2010; Dressing et al. 2010, Forgan 2012, 2014). With seasonal and latitudinal EBMs it is possible to calculate the fraction of habitable planet surface (Spiegel et al. 2008). In order to treat the vertical transport in a physical way one can couple EBMs with results from 1D climate calculations (Williams and Kasting 1997). Using this last approach, we have studied the influence of surface atmospheric pressure on liquid water habitability (Vladilo et al. 2013). To improve the modelling, we have then developed an Earth-like planet surface temperature model (ESTM) with a physically-based treatment of the horizontal transport (Vladilo et al. 2015). Here we apply the ESTM to discuss biological constraints of habitability different from the liquid water criterion. The main motivation for introducing new criteria of habitability is that the temperature limits provided by the liquid water criterion are not always representative of the thermal limits of life. For instance, some organisms (termed extremophiles) can live outside the liquid water temperature range, while the thermal limits of multicellular organisms on earth are often narrower than those of liquid water (see e.g. Clarke 2014). Here we discuss the possibility of defining limits of habitability for complex life. Limits for complex life are relevant for studies of Galactic habitability (Gonzalez et al. 2001, Lineweaver et al. 2004, Prantzos 2008, Gowanlock et al. 2011, Carigi et al. 2013, Spitoni et al. 2014). Temperature-based criteria of habitability for humans have been considered by Dole (1964). Here we introduce an index of habitability based on the thermal limits of multicellular life on earth. We then discuss whether such limits may also be appropriate for other forms of multicellular chemical life. An additional goal of our study is to investigate if, in addition to the ambient temperature, other planetary quantities accounted for by the climate model may help in setting biological constraints of habitability. In particular, we explore the role of the atmospheric columnar mass, which acts to shield the planet surface from ionizing radiation of astrophysical origin. As a word of caution, we recall that the climate model that we use does not track the chemical or biochemical evolution of the atmosphere, at variance with the approach of coupled geosphere-biosphere dynamics (Watson and Lovelock 1983, Wood et al 2008, Cresto Aleina et al 2013). Extension to such cases is deferred to future work.

**Definitions**

In this section we provide operational definitions of life, complex life and habitability used in this work. The definitions are designed in such a way that the environment should be able to host a biosphere with the following characteristics (1) detectable through observations of atmospheric biosignatures and (2) able to host complex life. The huge variety of terrestrial life forms and evolutionary pathways are used as a reference to infer general properties that may be shared by terrestrial and non-terrestrial organisms yet to be discovered, if any.

*Life*

The definition of life is the subject of intense debate in the literature (see e.g., Cleland and Chyba 2002, Kolb 2007, Lazcano 2008). For the purpose of our discussion we consider *a series of requirements* that should characterize, *by definition*, any form of life in the universe, focusing on properties relevant for the cross-feedback between life and its environment. With this pragmatic approach, we define life, terrestrial and non-terrestrial, with the following series of requirements.

At the most basic level, life is an active network of chemical reactions, the metabolism, leading to the synthesis and/or cleavage of functional molecular structures organized in cells. In conjunction with covalent bonding, hydrogen bonding is essential for building up any possible molecular structure of biochemical interest (see, e.g., Jeffrey and Saenger, 1991). Biochemical processes inside cells require a liquid medium/solvent; owing to the special properties of the water molecule (see, e.g., Bartik et al. 2011), we restrict our attention to water-based life, although we acknowledge the possibility of other chemistries that could form the basis of life under some circumstances (Schulze-Makuch and Irwin 2008, Budisa and Schulze-Makuch 2014). Cells must be able to grow and reproduce by means of a set of instructions, the genetic information, recorded in molecular form. At higher level of internal organization, complex life is composed of a coordinated network of differentiated cells specialized in different functions. Life inevitably undergoes Darwinian evolution[4], a gradual modification of the genetic instructions transmitted during reproduction, driven by long-term adaptation to ambient conditions.

At variance with the life that we know, non-terrestrial life could feature a different set of molecular structures and/or biochemical processes, and/or a different way of organizing/transmitting the genetic information, and/or any other possible difference at a higher level of complexity, as long as such differences are consistent with the above defining requirements.

---

[4] The existence of Darwinian evolution is generally considered to be an essential feature of life (Lazcano 2008).

*Complex life*

Analysis of the huge record of terrestrial evolutionary pathways indicates that unicellular organisms are not able to exploit all the potential resources offered by the environment. Multicellularity evolved many times on Earth, possibly as a response to environmental conditions (e.g. Grosberg and Strathmann 2007). By analogy, we assume that, given sufficient time and appropriate ambient conditions, multicellularity would also emerge outside Earth. Several lines of evidence indicate that oxygen metabolism is an essential requirement of complex life (Catling et al. 2005). The general lack of anaerobic metazoans (see however Danovaro et al. 2010) is probably due to the fact that only the oxygenic metabolism is sufficiently efficient to satisfy the energetic requirements that maintain the functional capabilities of complex life. There are some other terminal electron acceptors that could, in principle, fulfil the role that oxygen plays in earth-based life, and that might allow non-oxygen-based complex life to evolve on other planets. However, the chemical characteristics and high availability of oxygen in the universe makes it the most likely basis for the metabolism of complex life (Catling et al. 2005). We therefore assume that oxygen metabolism is essential for any form of complex life. We refer to Lane (2014) for a discussion on the bioenergetic constraints on the evolution of complex life.

Since this work is focused on the relationship between ambient temperature and life, we focus our attention to organisms whose internal temperature depends directly on and varies with ambient temperature. In terrestrial life these organisms are called *poikilotherms* because they feature a broad range of internal temperatures (e.g. Precht et al. 1973). Since we are interested in complex life, we focus on multicellular poikilotherms, such as plants (although note that a few plants have a limited form of endothermy, e.g. Minorsky 2003), invertebrates and ectothermic vertebrates. We omit homeotherms from our discussion because homeotherms are able to maintain a narrow range of internal temperature over a broad range of ambient temperatures (e.g. Ruben 1995) and this capability makes hard to establish thermal limits of habitability. Poikilotherms represent a necessary step along the

Darwinian pathway that has led to the emergence of homeotherms on Earth. Since homeotherms must evolve from poikilotherms, a planet must have some regions that are within the range of habitability for poikilotherms for homeothermy to emerge. In this sense, the limits of planetary habitability for poikilotherms are relevant for complex life in general, including homeotherms. This conclusion is reinforced by the fact that poikilotherms, such as plants, lie at the base of the food chain of homeotherms. Homeotherms have emerged only relatively late, a few 100 Myr ago (Hillenius and Ruben 2004), while multicellular poikilotherms have been present for at least 1 Gyr (Grosberg and Strathmann 2007). If these evolutionary time scales are sufficiently general, multicellular poikilotherms could be the most representative form of complex life in exoplanets.

The multi-path development of multicellularity on Earth had first to undergo prokaryotic to eukaryotic cell evolution, which may have taken place only once in Earth history, possibly as a stochastic endosymbiosis between two prokaryotes (e.g. Lane 2014, Koonin 2015). Recently, an archaeal phylum has been found to show traits that may have facilitated the symbiosis (Spang et al. 2015). The modality of the transition to eukaryotes affects the probability and timescale of subsequent multicellular development. We assume that the minimum timescale is driven by the oxygenation of the atmosphere (Section "*Habitability*", point (ii)).

In discussing the relationship between ambient temperature and habitability we need to exclude organisms with the technological capability of creating habitable conditions in an otherwise hostile environment. In any case, these organisms could not evolve in hostile conditions, because their evolutionary history would have included forms that did not have such technological capability. Therefore, the limits of habitability that we discuss are also relevant for the potential emergence of any technically able organism.

*Habitability*

Here we summarize the conceptual framework used to specify the concept of habitability. It is worth emphasizing that the conditions of habitability do not guarantee, per se, the emergence of life in a given planet (see e.g. Cleaves and Chalmers 2004).

*(i)* Surface habitability

We restrict our attention on the *surface* (or near surface) habitability of the planet. This choice is dictated by two considerations. First, surface life has the highest chance to produce atmospheric biosignatures: only in some special cases, detectable planetary features could provide an indirect link to subsurface biological activity (Parnell et al. 2010, Hegde and Kaltenegger 2013). Second, surface conditions required for habitability can be modeled with climate simulations that, eventually, will be tested with remote observations of the planet atmosphere.

*(ii)* Long term habitability

We restrict our attention on *long term* habitability conditions, i.e. conditions that persist for a time scale, $\tau_{hab} > \tau_{cxl}$, where $\tau_{cxl}$ is the time scale required for the development of multicellular life. This time scale can be constrained by the minimum level of oxygenation of the atmosphere advocated as a possible necessary precursor for complex life (Catling et al. 2005). Here we adopt a representative value $\tau_{cxl} \sim 2$ Gyr. The long term persistence of habitability is also a necessary condition for the persistence of a biosphere detectable during a significant fraction of the stellar life time. The persistence of habitability requires stable planetary orbits and a steady stellar luminosity. The constancy of stellar luminosity constrains the mass of the central star, since the time scales of stellar evolution become smaller with increasing stellar mass. The longest and most stable phase of stellar evolution is the "main sequence" phase of hydrogen burning. Based on the above considerations, this phase should last for a time $\tau_* \geq \tau_{hab} > \tau_{cxl} \sim 2$ Gyr. This yields an upper limit $M_* < 1.6$ $M_{sun}$ (Salaris and Cassisi, 2005).

Even during the main sequence the stellar luminosity gradually rises, affecting the climate and habitability. Therefore, the long term habitability requires a mechanism of climate stabilization that compensates for the increase of stellar luminosity. In the case of the Earth, the carbonate-silicate cycle has been proposed as a mechanism of long-term stabilization via negative $CO_2$ feedback (Walker et al. 1981). Such mechanism is essential in the definition of the classic habitable zone (Kasting et al. 1993; Kasting and Catling 2003). However, only planets with appropriate geophysical conditions

(e.g. tectonic activity) can maintain the carbonate-silicate cycle. Proving that such cycle is at work in exoplanets is a hard observational task. This possibility is currently investigated through modelling of planet interiors (e.g. Valencia et al. 2007, Stein et al. 2011, Frank et al. 2014). Alternatively, it may be the presence of life itself that compensates for the changes in the stellar luminosity, by modifying the atmospheric composition (Watson and Lovelock 1983).

*(iii)* Active metabolism

We make a distinction between life with active metabolism and organisms that can survive without growth and reproduction (e.g., Stan-Lotter 2007). We restrict our attention to *environments that allow organisms to have an active metabolism and to complete their life cycle* (Clarke 2014). An active metabolism is required to create the chemical disequilibrium that defines atmospheric biosignatures. The completion of the life cycle is required for the long-term maintenance of life, which in turn is required to provide enough time to darwinian evolution to originate complex forms of life.

*(iv)* Energy sources

We assume that energy sources for powering the metabolism are available at the planet surface. The requirement of energy sources can be translated in terms of power units (e.g., Shock and Holland 2007). Energy for life can be harvested from stellar insolation or chemical redox couples generated by geochemical processes. The level of stellar insolation typical of a planet in the classic habitable zone is in the order of $\sim 10^3$ W/m$^2$. This insolation is orders of magnitude higher than the minimum level required for photosynthesis (see, e.g. McKay 2014). Since the thermal limits of life that we discuss below are comparable to those dictated by the liquid water temperature range, the requirement of energy sources is automatically fulfilled, even in absence of geochemical sources of energy.

*(v)* Water

The minimum requirement of water supply is very low, since terrestrial life is also present in relatively dry environments (McKay 2014, Stevenson et al. 2015), although the capacity to survive in dry habitats is likely to be an evolutionary derived condition. Observations of protoplanetary disks will soon cast light on the process of water delivery during the stage of planet formation (e.g. Podio et al. 2013). Eventually, water absorptions will be searched in the atmospheric spectra of individual super Earths and possibly true Earth analogs (e.g. Seager and Deming 2010). In our climate calculations we assume that water is present in planet, the surface coverage of oceans being a free parameter of the model.

Long term habitability requires the *preservation* of water on the planet. Water can be lost as a consequence of a runaway greenhouse instability (Hart 1978, Kasting 1988). The climate conditions that determine the onset of the runaway greenhouse instability are used to locate the inner edge of the classic habitable zone (Kasting 1988, Kasting et al. 1993, von Paris et al. 2013, Kopparapu et al. 2013, 2014). Studies performed with moist, 3D climate models highlight the critical issues of these calculations, such as the uncertain treatment of the microphysics of the clouds (e.g. Leconte et al. 2013). In the ESTM we use an approximate estimate of the runaway greenhouse limit (see Vladilo et al. 2015) for the sake of comparison with previous work.

**Thermal limits of terrestrial life**

The limits of thermal tolerance of terrestrial life have been long investigated in the framework of studies of biophysical ecology (e.g. Gates 1980) and extremophiles (e.g. Cavicchioli and Thomas 2003). However, the experimental data on the temperature limits for the completion of the life cycle, $T_L$, are not particularly abundant, especially for complex life, and are spread in a large number of papers (see, e.g., Precht 1973). Here we use a recent review work by Clarke (2014) as the main reference for such data. In discussing the thermal limits we should always keep in mind that temperature is only one of the ambient factors that affect life. Changes in physiology, morphology, reproduction, movement, feeding rate, and so on, depend upon a set of ambient factors including humidity, wind, light intensity, and concentrations of oxygen and carbon dioxide, as well as past history (Gates 1980).

*Thermal limits for multicellular poikilotherms*

The thermal limits for terrestrial plants, invertebrates and ectothermic vertebrates known at the present time can be summarized as follows.

Lower limits for invertebrates and ectothermic vertebrates lie between ~ −2°C and ~0°C (Clarke 2014). A general figure adopted as the high temperature limit of ectothermic metazoans is 47°C (Pörtner 2002). For instance, the average upper limit for insects is evaluated as 47.4°C (Addo-Bediaco et al. 2000) and the limit for freshwater invertebrates at ~46°C (Clarke 2014). Also terrestrial and marine invertebrates generally conform to these limits. Possible exceptions are represented by a few species of nematodes that tolerate a temporary temperature peak of 60°C (Steel et al. 2013). This is line with the result that metazoans may tolerate temperatures beyond 47°C, but only for a short time (Schmidt-Nielsen 1997). Previous claims of much higher thermal limits of a marine invertebrate, *Alvinella Pompeiana*, have been dismissed by a recent study which indicates an upper limit ~50°C (Ravaux et al. 2013). As far as plants are concerned, the lower thermal limits of angiosperms are similar to those of ectothermic animals, $T_L \geq 0$°C, but the tolerance at high temperature is higher, with $T_L \lesssim$~ 65°C (Clarke 2014). The higher limit is conservative, since the thermal tolerance of *Dichantelium lanuginosum*, the plant that holds the high temperature record, is mediated by a mutualistic fungus and a mycovirus. In absence of such mediation, the plant can only grow up to 38°C (Márquez et al. 2007). Considering the uncertainties in these type of determinations, we adopt a representative interval 0°C $\leq T_L \leq$ 50°C for multicellular poikilotherms. As we discuss below (Section **Implications for the generation of atmospheric** biosignatures), these limits are consistent with one of the requirements of the definition of exoplanet habitability, namely the potential generation of biosignatures in the planetary atmosphere.

**Universal mechanisms of thermal response**

The temperature interval 0°C $\leq T_L \leq$ 50°C brackets the limits for growth and reproduction of *terrestrial* multicellular poikilotherms. In lack of experimental evidence of exobiology, we wonder whether a similar temperature interval may also apply to non-terrestrial life with similar characteristics (i.e., multicellular and without thermal control). To discuss this problem we recall some key aspects of the mechanisms of thermal response at different levels of biological complexity. Casting light on such mechanisms may reveal to which extent we may consider them to be universal, rather than specific of terrestrial life. Mechanisms of thermal response are most commonly investigated in the framework of studies of extremophiles (e.g. Cavicchioli and Thomas 2003), rather than in multicellular life. In recent times there is renewed interest in understanding the thermal response of complex life due to the potential impact of Earth's global warming on the geographical distribution of plants and metazoans (see, e.g., Pörtner 2002; Schulte 2015, Vasseur et al. 2015).

*Thermal performance curves*

The rates of metabolism, growth, development, or other biological rates, offer a way to measure the thermal performance of an individual organism, or a population, as a function of ambient temperature. The thermal performance curves (TPCs) obtained with these methods share a *common behavior*, well described by curves that rise up to a maximum at some optimal temperature and then fall steeply once this optimal temperature is surpassed (Schulte 2015, Vasseur et al. 2015). In spite of the presence of differences in the slopes of the TPCs versus temperature, or location of the optimal temperature, the same general shape of TPCs is found at different levels of biological complexity, including molecular structures, cells, organisms and populations. The same behavior is shared by different terrestrial species, often developed along independent evolutionary pathways. This common behavior supports the existence of universal mechanisms of thermal response. We consider the existence of such mechanisms at different levels of complexity, starting from the molecular level.

*Thermal response at the molecular level*

Mechanisms of thermal response are well known at the atomic and molecular level. For instance, the polarity of water decreases as the temperature increases, leading to a decline of the dielectric constant (Carr et al. 2011). This behavior affects the interaction of water with dissolved biomolecules, in particular lipids, but also proteins and nucleic acids (McKay 2014).

A key aspect of the mechanisms of thermal response is the kinetic energy of atoms/molecules, which is governed by the temperature, $T$. The initial rise of TPC can be understood in terms of the kinetic energy of the reactants taking part to biochemical processes. The observed behavior often obeys to the Arrhenius law, $k \sim e^{-(E_a/T)}$, where $k$ is the rate of reaction and $E_a$ the activation energy (Arrhenius 1915). The higher the temperature, the higher the reaction rate because the reactants are more likely to collide with enough energy to allow the reaction to occur. However, at some point the rise of kinetic energy would tend to randomize molecular structures by violent motion, leading to denaturation of the molecular structures. At this point, a rise of temperature would lead to a fast decline of the thermal performance. The gradual rise and fast decline of the TPCs are in qualitative agreement with such expectations, even though the detailed shapes require more complex explanations (Schulte 2015). It is logical to assume that the same behavior is also shared by non-terrestrial life of the type we have defined, i.e. life based on chemical processes taking place between molecular structures (see "Definitions"). Of course, the quantitative thresholds of temperature will depend on the particular molecular structures developed by different forms of life. Even so, broad quantitative thresholds with general validity can be inferred by studying the binding energies of biomolecules.

(i) Binding energy of biomolecules
Biomolecules are held together by weak, non-covalent bonds, such as hydrogen bonds and van der Waals interactions. Typical energies of hydrogen bonds lie in the range from 4 to 20 kJ $mol^{-1}$, while van der Waals interactions between 2 and 4 kJ $mol^{-1}$ (Berg et al. 2007). The interval $0^o C \leq T_L \leq 50^o C$ is centered around ~300 K, which corresponds to a mean kinetic energy $E_{kin} = 3/2 \, k \, T = 0.039$ eV, or ~ 4 kJ $mol^{-1}$, in terms of binding energy per molecule. If we gradually rise the temperature above ~ 300 K, the thermal energy will start to overcome the binding energy, leading to the denaturation of the biomolecules. Assuming that intramolecular and intermolecular hydrogen bonding is essential in any possible molecular structure relevant for life (see e.g. Jeffrey and Saenger, 1991), the thermal limits of such bonds suggest that the performance may start to decline approximately after ~ 300 K for any form of chemical life, not just the terrestrial one. To be more specific, we briefly discuss two examples of molecular structures relevant to this discussion, namely enzymes and molecular motors.

(ii) Thermal limits of enzymes
Biochemical reactions require the presence of catalysts to overcome their energy barriers.

Terrestrial life has developed enzymes, molecular structures that work as the catalysts of biochemical reactions. The efficiency of enzymes is temperature dependent[5] and the denaturation of enzymes sets an upper limit to the thermal performance of life, even though accounting for the detailed shapes of the temperature performance of protein activity requires models more complex than the classical two-state model based on kinetic rise and denaturation decline (e.g. Daniel et al. 2007, Lee et al. 2007, Peterson et al. 2007, Daniel and Danson 2010, Hobbs et al. 2013).

Molecular catalysts must emerge in any form of life based on chemical reactions, even though non-terrestrial life may develop enzymes with different molecular structure and different temperature of denaturation. We note that the considerations on the enzymes are relevant also for the photosynthesis: although the light dependent reactions of photosynthesis are mostly dependent on the photon flux, the light independent reactions, catalyzed by enzymes, are temperature dependent. At high temperature, while the gross gain extends up to ~70°C for a few species of cyanobacteria living in geothermal habitats (Ward and Castenholz 2000), the net gain of photosynthesis drops to zero between ~ 40°C and 50°C for a large variety of terrestrial plants (Precht et al. 1973, pp. 103-107)

(iii) Random thermal motion and molecular motors
Thermal energy is an efficient source of power at the nanometer scale typical of biological macromolecules. Many of such molecular structures are small enough to be shoved about by Brownian motion (i.e., random thermal motion) of the surrounding atoms and molecules, which occurs very rapidly at the nanometer scale. Cells possess molecular structures, called motor proteins that are able to convert random thermal fluctuations into a directed force (Fox and Choi 2001, Bustamante et al. 2001, Oster and Wang 2003). The temperature dependence of thermal motions suggests that the performance of molecular motors is confined to a restricted temperature range. An initial temperature rise would increase the Brownian motion and the efficiency of molecular motors. When the kinetic energy exceeds the binding energy of the molecules, a rise of temperature would make molecular motors less efficient, up to the point at which their molecular structure denatures. Molecular motors are ubiquitous (e.g. Berg et al. 2007) and essential in several aspects of life. The rise of their efficiency with increasing thermal energy provides a further evidence of temperature dependence at microscopic level, in addition to that of the chemical reactions mentioned above. Also in this case, the mechanism is universal: a temperature rise would increase, in any case, the Brownian motion from which a directed force can be extracted. The functioning of molecular motors sets a lower bound to the temperature compatible with life. In fact, in order to shove the molecular motor, the small molecules of the surrounding medium must be in a dynamic state, rather than in a rigid, crystal-like configuration. Since we are considering a water-based medium in our definition of life, this requirement implies that the water molecules surrounding the molecular motor must be in liquid phase, rather than icy phase. In our opinion, this consideration provides a new argument in support of the liquid water criterion, in addition of the requirement for a liquid medium able to dissolve and transport molecules inside the cell. This argument lends support the lower thermal bound $T \geq 0$°C for surface life, even though variations in the salinity of the medium would allow a slight decrease below ~0°C and supercooling (Precht 1973) would allow even lower temperatures, albeit in a thermodynamically unstable state.

*Thermal response at the unicellular level*

Unicellular organisms on Earth, and in particular prokaryotes, have developed several strategies of thermal adaptation. The upper threshold for prokaryotes can be as high as 122°C (*Methanopyrus kandleri*), while it is ~60°C (and possibly 70°C) for unicellular eukaryotes, such as algae and yeasts. The low limit of $T_L$ can be as low as −20°C for unicellular prokaryotes and eukaryotes. Broadly speaking, the strategies of high temperature adaptation of thermophilic prokaryotes rely on relatively rigid membranes and proteins, while low temperature adaptation of cryophilic prokaryotes are based on relatively loose, more flexible, membranes and proteins (Cavicchioli and Thomas 2003). Extant extremophiles use only one of such strategies at a time, i.e. either increasing or decreasing the molecular rigidity, even though the simultaneous presence of both strategies of adaptation is not impossible on biophysical grounds (Schulte 2015). The rigidity and flexibility of molecular structures can be changed, for instance, by varying the proportion of hydrophobic cores, hydrogen bonds, and thermophilic amino acids (Cavicchioli and Thomas 2003). One may wonder why multicellular life does not benefit of the broader thermal thresholds developed in the framework of the unicellular world. A possible explanation is that the changes of rigidity at the molecular level, with respect to some optimal value, might hinder the integration of higher level structures. For instance, changes of membrane fluidity can affect the functioning of proteins embedded in the mitochondrial membranes and can also diminish the rate of ATP production due to leakage of protons through the membrane (Schulte 2015). Another possibility, that we now discuss, is that some of the functions required to integrate processes across cells in multicellular organisms may have a lower maximum temperature threshold than the cellular and subcellular functions of unicellular life.

*Thermal response at the multicellular level*

(i) Thermal limits and organism complexity
A general result of the terrestrial data is that the thermal limits become narrower as the complexity of the organism, i.e. the number of sub-structures and functions, increases. For instance, among unicellular organisms, eukaryotic cells are much more complex than prokaryotic ones, and the $T_L$ interval is narrower for eukaryotes than for prokaryotes. Among eukaryotes, the $T_L$ interval is narrower for multicellular than unicellular organisms. A logical interpretation for this narrowing of thermal tolerance is that the integration of a large number of substructures and functions, which characterizes the rise of complexity, introduces new possibilities for limiting thermal resilience. For instance, in prokaryotes the thermal tolerance is only limited at the cellular and molecular level. Unicellular eukaryotes will experience, in addition, limitations deriving from compartmental coordination and organellar membranes. Multicellular life will experience, in addition, limitations resulting from central functions and coordination (see e.g. Fig. 2 in Pörtner 2002). Since the increase of structures and functions is exactly what defines the rise of complexity, the narrowing of thermal tolerance with increasing organism complexity is likely to be a common principle of life, terrestrial or not. Terrestrial homeotherms do not violate this principle, even though they break the link between internal and external temperatures. In fact, terrestrial homeotherms have evolved to maintain an internal temperature just below the highest ambient temperatures that poikilotherms can tolerate. This suggests that re-setting the maximum temperature for the complex interacting functions of a multicellular poikilotherm may be evolutionarily difficult.

(ii) Thermal mechanisms in multicellular eukaryotes
The large number of structures and functions present in multicellular organisms makes difficult to identify which mechanism may most effectively limit their thermal tolerance (Schulte 2015). The functions of the *oxygen delivery system* have been recently proposed as the first limit of thermal tolerance for ectothermic metazoans (Pörtner 2002). According to this interpretation, the thermal limits would be linked with an adjustment of the aerobic scope, i.e. the ratio of the maximal metabolic rate to the basal metabolic rate. The disturbances at the high systemic level of oxygen delivery would then transfer to lowest hierarchical levels, causing cellular and molecular disturbances. This scenario is in line with the expectation, discussed above, that the range of thermal tolerance decreases with increasing complexity of the organism. Several lines of evidence indicate that oxygen metabolism is an essential requirement of complex life (Catling et al. 2005). Since the oxygen delivery system is critically dependent on temperature (Schulte 2015), aerobic life is likely to be characterized by a limited thermal tolerance. Similarly, the nervous system, which coordinates functions across multicellular animals on earth, also fails at high temperatures. The presence of a nervous system and complex mechanisms for transporting oxygen may account for the slightly lower maximum tolerated temperatures of animals compared to plants (which lack both of these systems).

*Synopsis*

A comprehensive mechanistic understanding of the effects of temperature on biological processes is still lacking, particularly at the higher levels of biological complexity (Schulte 2015). In spite of this, the intimate nature of the main mechanisms of thermal response at work in terrestrial life suggests that the same mechanisms would also be at work in other forms of chemical life outside Earth, if any.

For water-based life, that we consider here, we argue that the water freezing point, $T \sim 0^{\circ}C$ for the range of pressures we consider in this work, is a universal lower bound because the kinetic energy of Brownian motion cannot be harvested by molecular motors below the water solidification point. The upper bound would be set by the temperature at which thermal energy denatures the molecular structure most sensitive to heat. Since the number of molecular structures potentially limiting the thermal tolerance must increase with increasing complexity of the organism, the thermal tolerance narrows as complexity increases. Oxygenic metabolism, apparently necessary for the development of complex life, sets tighter thermal limits than other high level functions typical of multicellular life. Taken together, the above conclusions strongly suggest that the thermal limits for complex life will be generally narrower than those implied by the liquid water interval usually adopted in studies of habitability. In lack of better indications, we adopt the thermal limits $0^{\circ}C \leq T_L \leq 50^{\circ}C$, representative of terrestrial multicellular poikilotherms, as a criterion of long-term habitability of complex life outside Earth. The fact that such limits are shared by multicellular poikilotherms emerged from independent evolutionary pathways on Earth is consistent with this assumption. As long as the evolutionary pathway from simple cells to complex multicellular organisms is a universal feature of life, the habitability requirements for multicellular poikilotherms can be taken as a precondition for the development of more advanced forms of complex life, such as homeotherms, for which it is harder to define thermal limits.

**Calculations of planetary habitability with climate models**

To implement a temperature-dependent index of surface habitability for exoplanets, we use the Earth-like surface temperature model (ESTM; Vladilo et al. 2015). This model features a set of climate tools that feed a seasonal and latitudinal EBM. The meridional transport is treated with a physically-based parametrization validated with 3D climate models. The vertical transport is described using radiative-convective atmospheric column calculations. The surface and cloud properties are described with algorithms calibrated with Earth experimental data and, whenever possible, based on physical recipes (e.g. the ocean albedo, which is a function of stellar zenith distance). The ESTM simulations iteratively search for a stationary solution of the local, instantaneous values of surface temperature, $T(\varphi, t_s)$, function of latitude, $\varphi$, and time, $t_s$. Here the time $t_s$ represents the seasonal evolution during one orbital period rather than the long-term evolution, which is not incorporated in the model. The matrix $T(\varphi, t_s)$ is a "snapshot" of the equilibrium surface temperature that can be calculated for different combinations of input planetary parameters. The parameters that can be varied include rotation rate, insolation, planet radius, gravitational acceleration, global cloud coverage, and surface pressure. In the present version of the ESTM the atmospheric content of non-condensable greenhouse gases (e.g. $CO_2$ and $CH_4$) can be varied as long as these gases remain in trace amounts in an otherwise Earth-like atmospheric composition. Long-term evolutionary effects can be simulated by assigning an evolutionary law to the parameter of interest. For instance, the impact of the luminosity evolution of the central star can be explored by using a suitable track of stellar evolution. The extremely low computational cost of the ESTM is ideal for exploring the impact of variations of a variety of parameters. More details on the model, and on the limits of validity of the input parameters, can be found in Vladilo et al. (2015).

To quantify the surface habitability we define a boxcar function $H(T)$ such that $H(T)=1$ when $T_1 \leq T(\varphi, t_s) \leq T_2$, and $H(T)=0$ when $T(\varphi, t_s) \notin [T_1, T_2]$, where $T_1$ and $T_2$ are the assigned temperature thresholds. We then integrate $H(T)$ in latitude and time. The integration in latitude is weighted according to the area of each latitude strip. The integration in time is performed over one orbital period. With this double integration we obtain a global and orbital average value of habitability, $h$, which represents the fraction of planet surface that satisfies the assigned temperature thresholds. We call $h_{lw}$ the index derived using the liquid water, pressure dependent temperature limits (Vladilo et al. 2013) and $h_{050}$ the index calculated by setting $[T_1, T_2]$ =$[0^{\circ}C, 50^{\circ}C]$. The index $h_{050}$ is representative of the habitability of complex life, in the sense that we have specified above. The use of the boxcar function $H(T)$ ignores the temperature dependence of the biological rates of complex life. Since the TPCs feature a slow rise above $\sim 0^{\circ}C$ and a fast decline after the optimal temperature (e.g.

Schulte 2015), $H(T)$ tends to overestimate the habitability in the proximity of the edges of the interval [0°C, 50°C]. As a result, the width of the habitable zone calculated with the index $h_{050}$ is somewhat larger than the effective width weighted according to the thermal response. In principle, one could account for the thermal response by replacing the boxcar function $H(T)$ with a normalized, temperature dependent TPC. Unfortunately, the exact functional dependence of the TPC varies between different species (Schulte 2015) and there is no single TPC representative of complex life as a whole.

**The atmospheric mass habitable zone**

The insolation that the planet receives from its central star, $S$ (Wm$^{-2}$), is one of the physical quantities that mostly affect the planetary climate and habitability. Thanks to this fact, the circumstellar habitable zone can be defined in term of an interval of distances from the central star that yields a suitable value of $S$. Obviously, the habitability is not only determined by $S$, but also by a variety of planetary quantities. In particular, we have shown that the surface atmospheric pressure, $p$, has a strong impact on the exoplanet surface temperature and habitability due to its influence on the vertical and horizontal energy transport (Vladilo 2013, 2015). Moreover, $p$ is directly linked with the atmospheric columnar mass, $N_{atm}=p/g$, which influences the habitability by shielding the planet surface from cosmic rays (see Appendix). Finally, $N_{atm}$ scales with the total atmospheric mass, $M_{atm} = 4\pi R^2 N_{atm}$, which is a diagnostics of the history of the planet (the atmospheric mass is the result of a complex history, starting with the accretion of volatiles at the epoch of planetary formation, and keeping on with the gas exchanges between the atmosphere, the surface/interior and the outer space at later epochs). By exploring the impact of $M_{atm}$ on the surface conditions one can in principle provide a link between the planet history and habitability. Given the importance of the atmospheric mass, here we explore variations of the habitability index $h$ in the plane ($S$-$N_{atm}$). In this way we build up maps of habitability in the plane ($S$-$N_{atm}$). Regions where $h > 0$ define a habitable zone that we call atmospheric mass habitable zone (AMHZ).

The AMHZ is a diagnostic tool of habitability that can be compared with experimental data of $S$ and

$N_{atm}$. The local, instantaneous insolation $S(\varphi,t_s)$ can be calculated from the exoplanet orbital parameters and the luminosity of the central star. The long term evolution of $S$ can be modelled with a suitable set of evolutionary tracks of stellar luminosity (e.g. Bressan et al. 2012). The atmospheric columnar mass $N_{atm}=p/g$ can be calculated given an estimate of $p$, since $g$ can be calculated from the planetary mass, $M$, and radius, $R$, obtained from classic observational methods of exoplanets (e.g. Mayor et al. 2014, Batalha 2013 and refs. therein). A rough estimate of $p$ can be obtained from the relation $p = gM_{atm}/(4\pi R^2)$, where $M_{atm}$ is estimated as the fraction of planet mass that is expected to be incorporated in the atmosphere (e.g., Kopparapu et al. 2014). Eventually, $p$ could be estimated from transmission spectra of planetary atmospheres carried out with next generation instruments, with the aid of specially designed spectroscopic tools (e.g. Misra et al. 2013). Alternatively, one could estimate the average $N_2$ column density, which is the main contributor to $N_{atm}$ in an Earth-like atmosphere, from spectroscopic signatures of the collisions between $O_2$ and $N_2$ molecules detectable in $O_2$-bearing atmospheres (Pallé et al. 2009).

In Figs. 1 and 2 we present a series of AMHZ maps, where the color coding scales with the value of the habitability, which is calculated using the index $h_{lw}$ (in the left panel of Fig. 1) and $h_{050}$ (in all the other cases). Each map is obtained from a large number of simulations covering a grid of $S$ and $N_{atm}$ values, while keeping fixed the remaining stellar, orbital and planetary parameters. Unless differently specified, Earth parameters are adopted while varying $N_{atm}$ and solar parameters while varying $S$. In the AMHZ maps the insolation increases from right to left for consistency with the classic HZ, where the distance from the star increases from left to right (e.g., Kasting et al. 1993). To compare our results with the classic inner edge of the HZ, we plot in each map an approximate estimate of the runaway greenhouse limit (red curve; based on the results of Leconte et al. 2013, we define the limit from the condition that the columnar mass of water vapour exceeds 1/10 of the total atmospheric columnar mass, see Vladilo et al. 2015). Locations to the left of this edge represent cases in which the planet can preserve its water only for a relatively short period of time, from the onset of the runaway greenhouse instability. At variance with the classic HZ, the atmospheric composition is kept fixed in each individual AMHZ. The impact of variations of atmospheric composition, or of other orbital/planetary parameters, can be investigated by building up the AMHZ for a different set of input parameters. In Fig. 2 we illustrate the results obtained by varying $pCO_2$ in trace amounts.

*Mean global temperature and habitability in the AMHZ*

The solid green curves in Figs. 1 and 2 indicate the isotherm lines where the mean global orbital temperature of the planet surface, $T_m$, is constant. The isotherms highlight the trend of $T_m$ with $S$ and $N_{atm}$. One can see that $T_m$ tends to rise both with increasing insolation $S$ and increasing $N_{atm}$ (i.e., increasing strength of the greenhouse effect). As a result, the isotherms are generally tilted. However, the slope of the isotherms changes in different regimes of $N_{atm}$. At high $N_{atm}$ the slopes are relatively large, whereas at low $N_{atm}$ they are generally small. This behavior is particularly clear for the isotherm $T_m = 50^oC$ (green curve close to the inner edge), for which the separation between these two regimes lies at $N_{atm} \sim 3 - 5 \times 10^2$ g/cm$^2$ (see Fig. 1). At high $N_{atm}$, the isotherm gets closer to the star if the atmospheric mass decreases. At low $N_{atm}$, the atmospheric radiative transport becomes negligible and a decrease of $N_{atm}$ does not help to keep habitable a planet with high insolation.

The color coding of Figs. 1 and 2 show that when the mean global temperature gets close to the assigned thermal thresholds, the mean fraction of planet surface that is habitable during an orbital period tends to vanish. The mean global orbital habitability is determined by the fraction of time that each latitude strip spends inside or outside the assigned temperature limits. In the examples shown in Fig. 5, the mean habitability $h_{050}$ depends on the zones reaching $T(\varphi) > 50^{\circ}C$ at low latitude (bottom panels) and the zones with $T(\varphi) < 0^{\circ}C$ at high latitude (top panels).

*"Liquid water" versus "complex life" habitable zone*

In Fig. 1 we compare the maps of habitability obtained for the liquid water index, $h_{lw}$ (left panel), and the complex life index, $h_{050}$ (right panel), for the same set of Earth-like parameters. The outer edge of habitability is identical in both cases because the lower temperature limit is $0^oC$ for both indices. The outer edge shifts to lower values of $S$ with increasing $N_{atm}$ because the greenhouse effect becomes stronger and the condition $T > 0^oC$ is satisfied at lower insolation. As far as the inner edge is concerned, the results obtained for the two indices are instead quite different. When $N_{atm}$ increases, the inner edge shifts to low $S$ for the $h_{050}$ index (right panel) and to high $S$ for the $h_{lw}$ index (left panel). In this latter case, based on the liquid water criterion, the behavior of the inner edge is due to the rise of the water boiling point with increasing $p$. However, the region to the left of the runaway greenhouse limit (red line) does not comply with the requirement of long term habitability. Instead, in the case of the complex life index $h_{050}$, most of the habitable zone lies to the right of runaway greenhouse limit, in line with the requirement of long term habitability.

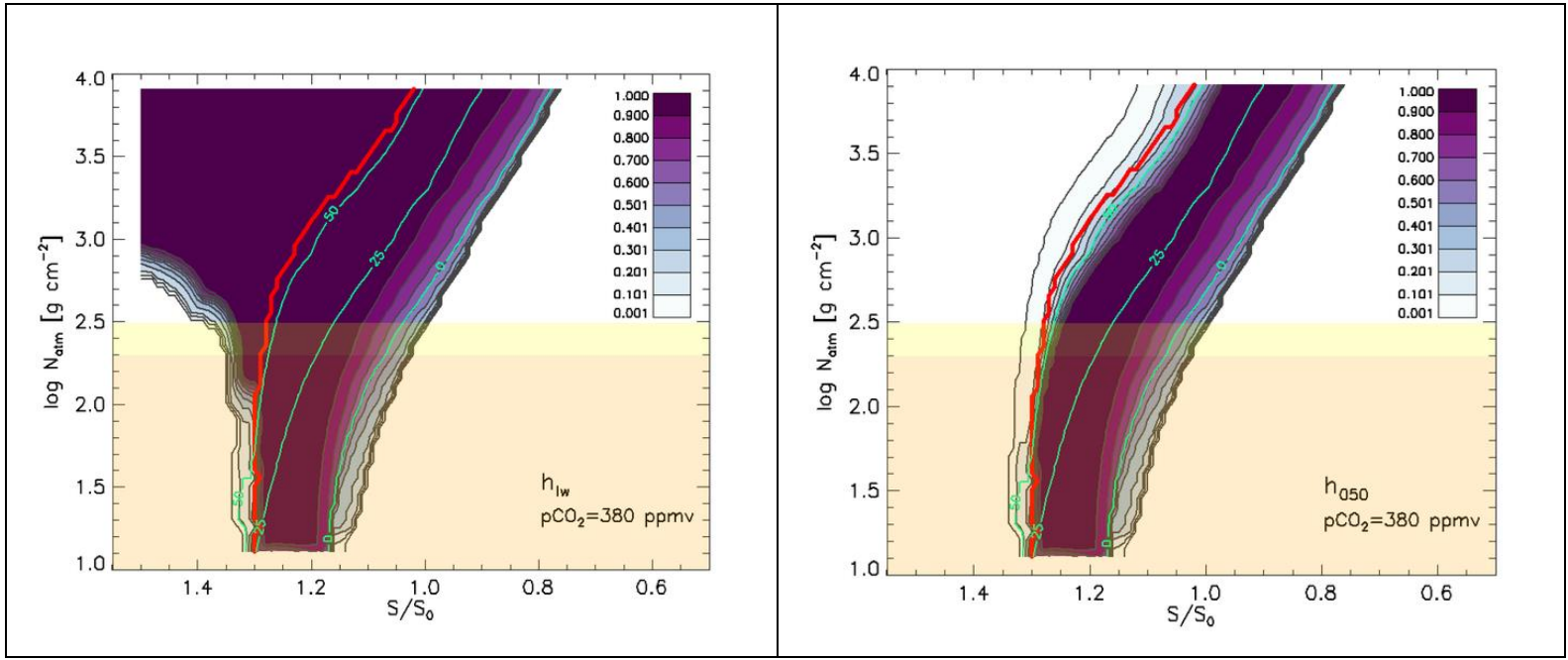

Figure 1- Atmospheric mass habitable zone (AMHZ) of an Earth like planet as a function of normalized insolation, $S_{eff}=S/S_o$ ($S_o=1360$ W m$^{-2}$), and atmospheric columnar mass, $N_{atm}$ (Earth value of $N_{atm}$ is 1033 g cm$^{-2}$. A constant Earth-like atmospheric composition is adopted, with p$CO_2$=380 ppmv. The color of the maps scale according to the habitability index $h_{lw}$ (left panel) and $h_{050}$ (right panel). Green curves: isothermal contours $T_m$ =0$^o$C, 25$^o$C and 50$^o$C. Red line: approximate estimate of the runaway greenhouse instability limit (see text). Shadowed regions: range of $N_{atm}$ values where the surface radiation dose of secondary particles of GCRs is >100 mSv/yr for a planet without magnetic field (yellow) and with an Earth-like magnetic moment (orange); see text and Appendix.

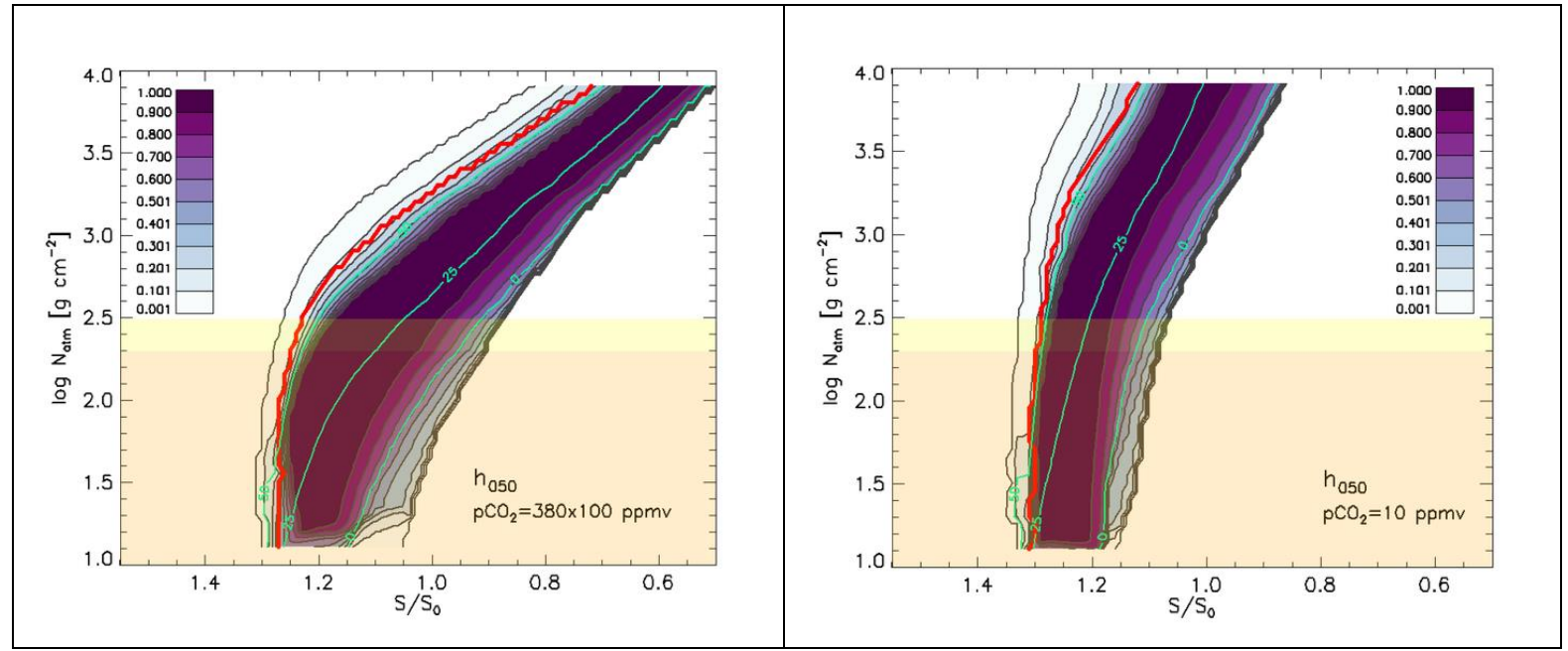

Figure 2 - AMHZ for the habitability index $h_{050}$, with a 100 fold rise (left panel) and 0.0263 fold decrease (right panel, corresponding to pCO2=10 ppmv) of the $CO_2$ partial pressure with respect to the Earth representative value pCO2=380 ppmv. Symbols, curves and shading as in Fig. 1.

*Comparison with the habitable zone of land-dominated planets*

Planets dominated by land, rather than oceans, may avoid the runaway greenhouse instability up to high values of insolation thanks to their low atmospheric water vapor content. For such water-limited ("dry") planets the inner edge of the habitable zone lies closer to the star than the classic inner edge (Abe et al. 2011; Zsom et al. 2013). Although the study of land-dominated planets is beyond the purpose of the present work, the comparison between the results obtained with the $h_{lw}$ and $h_{050}$ indices illustrates how biological limits prevent the presence of complex life at high values of insolation even in absence of the runaway greenhouse instability. Indeed, Figs. 1 and 3 indicate that the $h_{050}$ inner edge for complex life is significantly more distant from the star than the $h_{lw}$ inner edge, which is representative of the boiling point in absence of the runaway greenhouse instability. The extended limits of inner edge of dry planets (Abe et al. 2011; Zsom et al. 2013) can be suitable for extremophiles, but not for complex life. It is worth mentioning that evapotranspiration from vegetation, if present, could trigger a hydrological cycle and favor the presence of a moist atmosphere also on a sandy planet without oceans (Cresto Aleina et al. 2013).

*Variations of atmospheric composition*

Fig. 2 illustrates the effect of variations of the $CO_2$ partial pressure on the habitability map calculated with the index $h_{050}$. In the left panel, $pCO_2$ is raised by a factor of 100 compared to the Earth's reference value $pCO_2$=380 ppmv, while in the right panel $pCO_2$ is decreased to get a value of 10 ppmv, representative of the minimum value for plants with C4 photosynthesis systems (Caldera and Kasting 1992). In the first case the whole AMHZ is displaced to lower $S$ values due to the higher strength of the greenhouse effect. The effect becomes stronger at high values of atmospheric columnar mass. In the case of low $pCO_2$ (right panel) the AMHZ is displaced to higher values of $S$. At this very low value of $pCO_2$ the slope of isotherms becomes almost negligible, and the temperature, habitability, position and width of the HZ are determined by $S$, being weakly dependent on $N_{atm}$.

Variations of insolation affect not only the temperature, but also the amount of water vapor, which changes at different locations inside AMHZ. By adopting a representative value of relative humidity, $r$=0.6, the ESTM calculates $p(H_2O)$ as a function of planet surface temperature. The water vapor pressure model is predicted to rise from the outer edge to the inner edge due to the temperature dependence of the water vapor saturation pressure. As an example, for an Earth-like planet with $N_{atm}$= 1033 g cm$^{-2}$ and $pCO_2$=380 ppmv, the ESTM predicts that the mean global $p(H_2O)$ increases from ~3.5 mbar at $T_m$= 0$^o$C ($S_{eff}$=0.934) to ~ 72 mbar at $T_m$= 50$^o$C ($S_{eff}$=1.187).

*Width of the habitable zone*

In Fig. 3 we plot the insolation limits of habitability, $S_{min}$ and $S_{max}$ (curves at the top of the figures) and the width of the AMHZ, $\Delta S$=$S_{max}$−$S_{min}$ (curves at the bottom) as a function of $N_{atm}$ for different model parameters. In the left panel we compare the results obtained with the liquid water criterion (dotted light-blue curves) and the complex life $h_{050}$ index (solid green curves) for a planet with Earth-like atmospheric composition. With the $h_{lw}$ index the inner edge $S_{max}$ and the width $\Delta S$ rise significantly when $N_{atm}$>200 g/cm$^2$. The complex life index $h_{050}$ yields instead a gradual decline of $S_{max}$ and a modest increase of $\Delta S$ with increasing $N_{atm}$. In the right panel we show the impact of variations of of $pCO_2$ on the insolation limits calculated with the index $h_{050}$. The limits of insolation tend to decrease with increasing $N_{atm}$. The variations become stronger as $pCO_2$ increases from 10 ppmv (blue, dashed-dotted line), to the reference Earth value 380 ppmv (green solid curve), and finally up to 38000 ppmv (red, dashed curve). Variations of the inner edge as a function of $N_{atm}$ are quite complex because the greenhouse effect in the proximity of the inner edge is governed not only by variations of $N_{atm}$ and $CO_2$, but also by water vapor, whose partial pressure is temperature dependent. Variations of the outer edge versus $N_{atm}$ are more regular, probably because water vapor plays a negligible role at low temperatures. At the bottom of Fig. 3 we plot the width of the AMHZ, $\Delta S$=$S_{max}$−$S_{min}$, for the same three values of $pCO_2$. In general, $\Delta S$ increases with increasing $pCO_2$. However, in the curve $p\mathrm{CO}_2 = 38000$ ppmv, the strong slope of the inner edge results in a decrease of $\Delta S$ at $N_{\mathrm{atm}} > 2000$ g/cm$^2$.

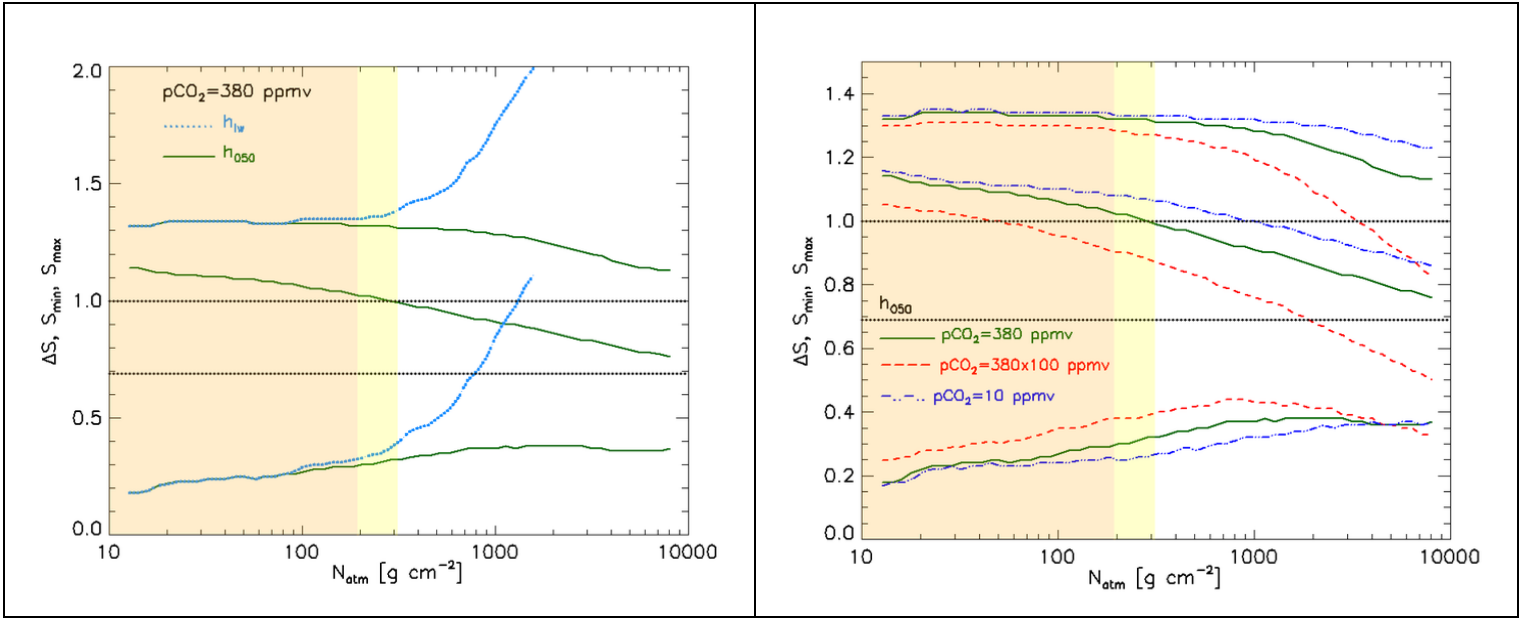

Figure 3- Insolation limits $S_{\mathrm{min}}$ and $S_{\mathrm{max}}$ (upper curves) and width $\Delta S = S_{\mathrm{max}} - S_{\mathrm{min}}$ (lower curves) of the AMHZ as a function of $N_{\mathrm{atm}}$ for different indices of habitability and levels of $p\mathrm{CO}_2$. Left: comparison of the results obtained using the $h_{\mathrm{lw}}$ (dotted light-blue curves, computed up to $S_{\mathrm{eff}}=2$) and $h_{050}$ (solid green curves) indices for a planet with $p\mathrm{CO}_2=380$ ppmv. The outer edge of $h_{\mathrm{lw}}$ is the same as for $h_{050}$. Right: effects of variations of $p\mathrm{CO}_2$ using the $h_{050}$ index. Blue dot-dashed curves: $p\mathrm{CO}_2=10$ ppmv. Green solid curves: Earth reference value $p\mathrm{CO}_2=380$ ppmv. Red dashed lines: $p\mathrm{CO}_2=38000$ ppmv. Black dotted lines: insolation S at the initial main sequence (bottom) and at 4.58 Gyr (top) for a planet at 1 AU around a sun like star.

## Ambient conditions and life inside the habitable zone

Changes of ambient conditions that depend upon $S$ and $N_{\mathrm{atm}}$ will determine which type of life is potentially present at different locations ($S$, $N_{\mathrm{atm}}$) inside the AMHZ. Here we show some examples that illustrate the influence of changes of temperature, surface radiation dose and atmospheric composition.

(i) Mean temperature and temperature excursions

The mean global orbital temperature $T_{\mathrm{m}}$ gives a broad indication of which type of life, classified according to its thermal tolerance, might be present at different locations in the plane ($S$, $N_{\mathrm{atm}}$). In the proximity of the inner and outer edge $T_{\mathrm{m}}$ is close to the upper and lower thermal limits of terrestrial multicellular poikilotherms, respectively. If such limits are universal, we may expect a dearth of poikilotherm species adapted to such extremes, while thermophilic and cryophilic extremophiles would find favorable conditions near the inner and outer edge, respectively. In the proximity of the outer edge, the effective habitability for poikilotherms would be low because their thermal performance rises slowly at $T \geq 0^{\mathrm{o}}$C. At high temperature the capability of hosting multicellular poikilotherms would depend on the optimal temperature of different species potentially present in the planet. For each species the habitability would drop steeply after the thermal optimum, owing to the fast decline of TPCs at $T > T_{\mathrm{opt}}$.

Besides the mean temperature $T_{\mathrm{m}}$, latitudinal and seasonal excursions of temperature will affect the distribution of different forms of life according to their thermal tolerance. To show these effects we calculate the mean orbital profiles of surface temperature as a function of latitude, $\underline{T}(\varphi)$, by integrating the temperature matrix $T(\varphi,t_s)$ during one orbital period. Similarly, we calculate the mean orbital habitability $\underline{H}_{050}(\varphi)$ by integrating $h_{050}$ during one orbital period. In the top panels of Fig. 4 we show $\underline{T}(\varphi)$ (left panels) and $\underline{H}_{050}(\varphi)$ (right panels) for a sequence of ESTM simulations selected along a line of constant atmospheric columnar mass, $N_{atm}=1033$ g/cm$^2$, inside the AMHZ of Fig. 1 (right panel). A similar example is shown in the lower panels, but by combining a decreasing amount of $CO_2$ for increasing $S$, the exact sequence of profiles depending on how $S$ and $f_{gh}$ are combined (see Section '*Time interval of habitability*'). In any case, in the proximity of the outer edge of habitability (blue solid curves) only an equatorial zone of the planet lies above the water freezing point (cyan horizontal line in the left panel), and the habitability $\underline{H}_{050}(\varphi)$ drops at high latitudes. In the proximity of the inner edge the equator is too hot and only the polar regions are habitable (black, dashed-dotted curves).

The mean orbital temperature-latitude profiles $\underline{T}(\varphi)$ inside the AMHZ can be steep or flat, depending on the location of the planet in the plane ($S$, $N_{atm}$). These latitude profiles become flatter with increasing $N_{atm}$ because the efficiency of the horizontal transport tends to rise with increasing $N_{atm}$ (e.g. V15). The profiles also tend to become flatter with increasing $S$, as in the examples of Fig. 4, because the latent heat component of the horizontal transport rises because $pH_2O$ increases at high temperature. For flat temperature profiles, life at different latitudes would show little diversity as far as the thermal response is concerned. Conversely, a large meridional gradient of temperature would favor the emergence of a diversity of life forms, each one adapted to the mean temperature of its own latitude zone. In each latitude strip, the temperature will vary around its mean orbital value $\underline{T}(\varphi)$ as a result of seasonal excursions.

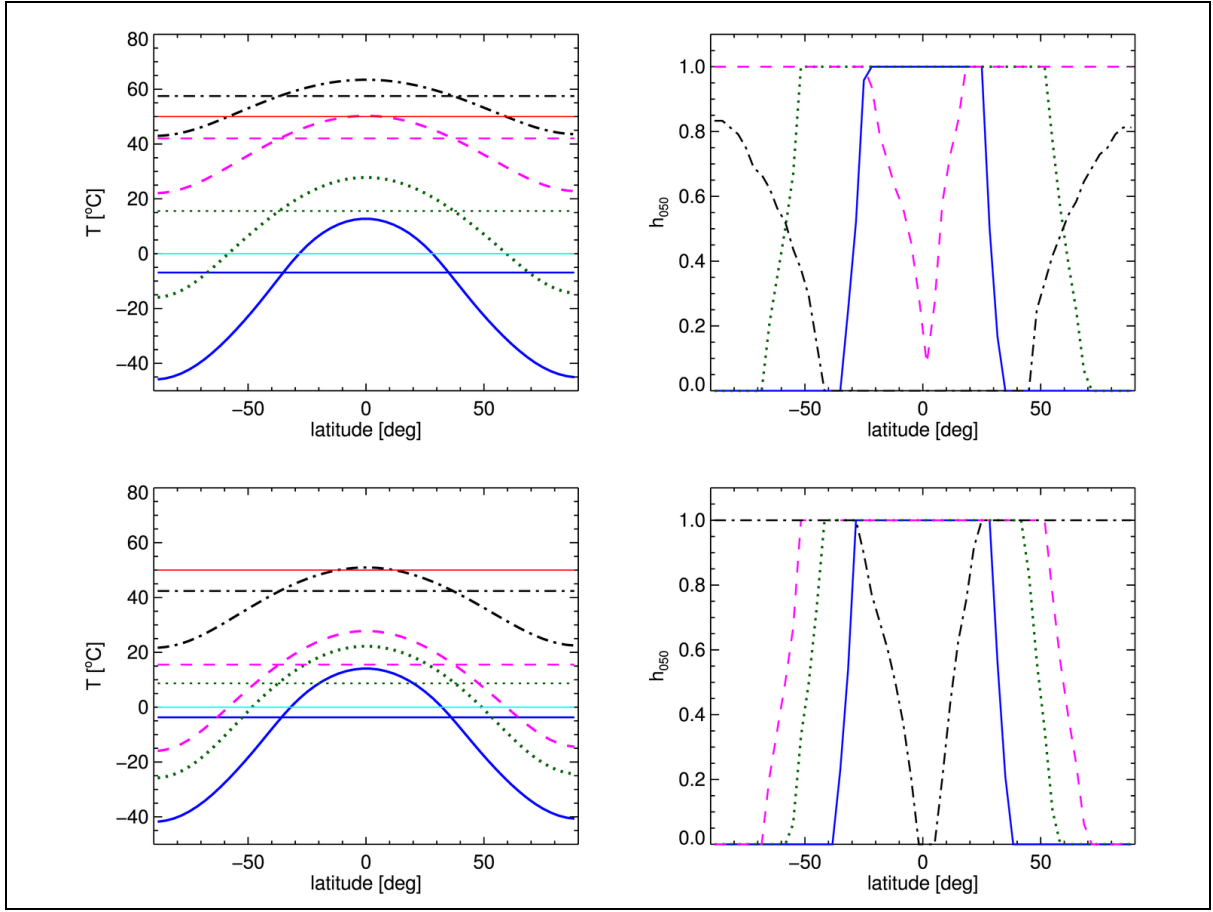

Figure 4 – Effects of variations of insolation on the mean orbital latitude profile of surface temperature, $\underline{T}(\varphi)$ (left panels), and habitability, $\underline{H}_{050}(\varphi)$ (right panels), for an Earth-like planet with atmospheric columnar mass, $N_{atm}$=1033 g/cm². Top panels: model predictions at constant $f_{GH}$=pCO2/(380 ppmv)=1 with insolation S=0.92, 1.00, 1.15 and 1.22 (blue solid, green dotted curves, dashed magenta, and black dashed-dotted curves, respectively). Bottom panels: predictions for a planet in which pCO2 decreases as the insolation increases. The blue, green, magenta, and black curves correspond to (S, $f_{GH}$) = (0.77,100), (0.90,10), (1.00,1), and (1.22,0.026), respectively. The mean orbital global temperature of each $\underline{T}(\varphi)$ profile is shown with an horizontal line of the same color.

Seasonal variations are induced by the eccentricity of the orbit, $e$, which can be measured in exoplanets, and by the tilt of the rotation axis, $\varepsilon$, which can be parametrized in the climate simulations. In Fig. 5 we show examples of the seasonal excursions of surface temperature of an Earth-like planet with 4 possible combinations of insolation and atmospheric columnar mass inside the AMHZ. In each panel the seasonal excursions are displayed by superposing instantaneous temperature-latitude profiles calculated at different phases of one orbital period (thin curves). One can see that the seasonal variability is negligible in the equatorial belt, but increases towards the poles. At a given latitude the seasonal excursions are larger at low $N_{atm}$ (left panels) than at high $N_{atm}$ (right panels) because the atmospheric columnar mass damps variations of surface temperature. Also variations of the axis tilt and geographical distribution of continents/oceans, not discussed here, would affect the temperature excursions. Variations of the axis tilt would affect the temperature excursions at high latitude and only modestly at low latitude.

Seasonal excursions affect the potential geographical distribution of different forms of life. Latitude zones with strong seasonal excursions would be expected to favor the emergence of organisms with capability of seasonal migration and/or organisms that can function at a wide range of different body temperatures, such as terrestrial eurytherms. Equatorial regions and, more in general, planets with high atmospheric mass would be expected to favor the emergence of organisms only capable of living in a narrow temperature range, such as terrestrial stenotherms. At the present time, the lack of understanding of the origin of terrestrial homeotherms prevents conclusions about the existence of a relationship between ambient temperature conditions and the emergence of homeothermy. Terrestrial homeotherms may have evolved in response to inter-specific competition (Lovegrove 2012), rather than in response to ambient conditions.

(ii) Surface radiation dose

The planet atmosphere shields the life potentially present on the planet surface from the biological damage induced by ionizing radiation of astrophysical origin (see Appendix). The shielding from cosmic rays can be used to assess the habitability of the planet according to the value of atmospheric columnar mass $N_{atm}$ and to the radio tolerance of different types of organisms. Based on the discussion in the Appendix, we adopt here 100 mSv/yr as an illustrative limit of long term irradiation tolerable by complex life. From the results obtained by Atri et al. (2103) for an Earth-like planetary magnetic moment, a surface radiation dose of 100 mSv/yr corresponds to an atmospheric columnar mass $N_{atm}$ ~200 g/cm$^2$. The shaded region in Figs. 1 and 2 indicate the range of low atmospheric column densities that yield a radiation dose in excess of 100 mSv/yr. The $N_{atm}$ threshold depends on the planetary magnetic moment, as shown in the figure: the orange and light yellow shaded areas indicate the thresholds obtained for an Earth-like magnetic field and for a planet without magnetic field, respectively. The results of such calculations are quite independent of the exact chemical composition of the atmosphere (Atri et al. 2013). These results indicate that in the shaded areas of low $N_{atm}$ of the AMHZ the high radiation dose may favor genetic mutations, accelerating the pace of darwinian evolution, or may lead to the reinforcement of biological mechanisms aimed at repairing the damages induced by ionizing radiation. The high radiation dose may combine with other ambient conditions discussed above, such as the latitudinal and/or seasonal temperature excursions, to produce a variety of possible combinations of ambient conditions that differentiate the type of habitability at different locations in the AMHZ. For instance, a planet with $N_{atm}$ ~ 400 g cm$^{-2}$ would experience, at the same time, high seasonal excursions of surface temperature (see above discussion of Fig. 5) and a relatively high surface dose of radiation, ~ 20-40 mSv/yr. These ambient conditions would favor the emergence of species with large capability of thermal adaptation and radio tolerance. Conversely, the right panels of Fig. 5 show examples with small temperature excursions and low radiation dose resulting from the relatively high atmospheric column density ($N_{atm}$= 5725 g cm$^{-2}$). These ambient conditions may favor the emergence of species without particular requirements in terms of radio tolerance and adapted to a narrow range of temperatures in each latitude strip. All together, this latter situation would probably be less dynamic, in terms of darwinian evolution, in comparison with the case of low $N_{atm}$ and high radiation dose discussed above.

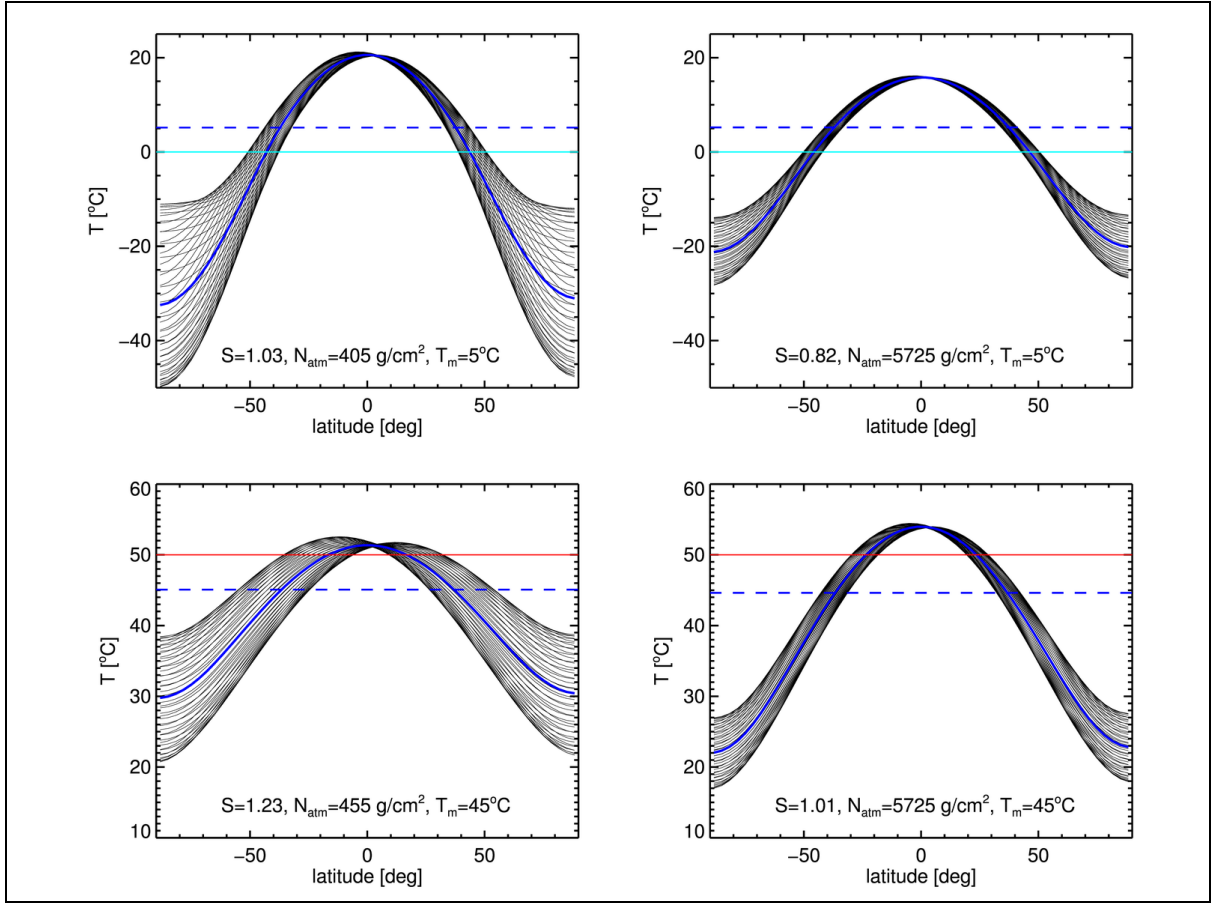

Figure 5 - Latitude-temperature profiles of an Earth-like planet in 4 representative locations of the AMHZ. Top panels: locations in the proximity of the outer edge, with mean global orbital temperature $T_m=5^{\circ}C$ (dashed line in upper panels). Bottom panels: locations in the proximity of the inner edge, with $T_m=45^{\circ}C$ (dashed line in lower panels). The values of insolation, S, and atmospheric columnar mass, $N_{atm}$, are indicated in each panel. Thin curves: seasonal evolution of $T(\varphi)$ during one orbital period. Thick curves: mean orbital temperature, $\underline{T}(\varphi)$. Adopted orbital eccentricity: e=0.0167; rotation axis tilt: $\varepsilon=22.5^{\circ}$.

(iii) Atmospheric composition and water vapor content

Variations of $p\mathrm{CO_2}$ affect the type of life potentially present in the planet. In particular, the photosynthesis requires $\mathrm{CO_2}$ and this fact can be used to set a lower threshold to the possible values of $p\mathrm{CO_2}$. On Earth, plants with C3 and C4 photosynthetic systems require a minimum $p\mathrm{CO_2}$ of ~150 and 10 ppmv, respectively, while the photosynthesis of (some) cyanobacteria requires a minimum $p\mathrm{CO_2}$ of 1 ppmv (Caldeira & Kasting 1992; O'Malley-James et al. 2012 and references therein). The lowest possible value for the plant photosynthesis, $p\mathrm{CO_2}=10$ ppmv, can be considered as a representative limit for efficient oxygenic production by means of photosynthesis. Since the presence of oxygen is required for the existence of aerobic metabolism and complex life, the inner edge of the AMHZ calculated at $p\mathrm{CO_2}=10$ ppmv (right panel of Fig. 2 and Fig. 3), is illustrative of the minimum inner edge (maximum insolation) for a planet hosting complex life.

Also variations of humidity affect life, in particular when they are associated with variations of temperature. Therefore the expected rise of water vapor with increasing temperature will vary the characteristics of habitability inside the AMHZ. In the proximity of the outer edge life should be adapted to cold, dry conditions, while in the proximity of the inner edge to damped, warm conditions.

## Habitability evolution

As a result of the stellar, orbital and planetary evolution, the planet will move in a plane ($S$, $N_{atm}$), possibly inside the AMHZ. Tracking the habitability evolution of the planet in the initial stages of stellar evolution and planetary formation is extremely difficult because such stages can be characterized by fast variations of stellar flux and orbital parameters. We therefore consider the situation in which the host star has entered the main sequence (MS) and the planetary system has attained dynamical stability. At this stage the luminosity evolution of the central star can be used to track the evolution of the planet insolation. Moreover, the accretion of volatiles is ended and, after the formation of a solid crust, variations of the atmospheric mass and composition become mild, as long as the planet is able to avoid atmospheric loss. In these conditions the planet is expected to cross the AMHZ at roughly constant $N_{atm}$ and increasing $S$, owing to the gradual rise of the star luminosity in the main sequence evolution. As a result of the gradual migration of the planet in the AMHZ, the type of life potentially present in the planet will vary as discussed in the previous section. For instance the geographical distribution of complex life will shift from low latitude to high latitude with increasing insolation. The evolution of habitability can be calculated using stellar evolutionary tracks. For illustrative purposes here we adopt a PARSEC evolutionary track (Bressan et al. 2012) tailored for the sun[6], with initial metallicity $Z_{ini}$=0.01774, helium abundance $Y_{ini}$=0.28, main sequence time $\tau$* = 11 Gyr, and current age 4.58 Gyr.

### Time interval of complex life habitability

As we discussed above, long term habitability is a necessary (even though not sufficient) condition to allow the long term existence of a biosphere and the ensuing evolution to complex life forms. The stellar evolutionary tracks provide a relation between the age of the star, $t$*, and its luminosity that we can use to calculate the time span of habitability. For instance, we can convert the range of insolation $\Delta S_{hab}$=$S_{max}$–$S_{min}$, calculated at constant $N_{atm}$ between the edges of the habitable zone, into a *maximum time of habitability*

$$\Delta t_{hab}(N_{atm}) = t*(S_{max}) - t*(S_{min})$$

With this definition, $\Delta t_{hab}$ is positive during the main sequence phase of hydrogen burning, when the stellar luminosity is characterized by a steady, gradual rise. The actual fraction of habitable surface will change with time, depending on the instantaneous value of $N_{atm}$ and $S_{eff}$. Therefore, a more meaningful quantity can be provided by weighting each time step with the corresponding value of mean habitability $h$ ($h_{050}$ or $h_{lw}$) at each location $i$ inside the AMHZ. In this way, we obtain an *effective time of habitability*

$$\tau_{hab}(N_{atm}) = \sum_i h_i \, \delta t_i*$$

where $\delta t_i$* is the time increment corresponding to a constant increment of insolation, $\delta S$. The increment $\delta t_i$* is not constant because it depends on the rate of luminosity evolution of the star, which changes with time. For instance, for our adopted sun track, the luminosity rate increases from ~0.05 $L_{sun}$/Gyr at the first stage on the MS, to the current ~0.08 $L_{sun}$/Gyr, and will reach ~0.35 $L_{sun}$/Gyr towards the end of the MS. Since the rate of growth of the luminosity increases with age, the early stages of MS will experience longer intervals of habitability $\tau_{hab}$ for a given interval $\Delta S_{hab}$ than the late stages of MS.

In Fig. 6 we show calculations of $\Delta t_{hab}$ and $\tau_{hab}$ performed at different values of $N_{atm}$ for 3 values of $CO_2$ in an otherwise Earth-like atmospheric composition. The planet is assumed to lie at 1 AU from the central, sun-like star. One can see that the times of habitability become larger with increasing $N_{atm}$ and increasing $pCO_2$. From this figure we can constrain the values of $N_{atm}$ and $pCO_2$ that yield an effective time of habitability compatible with our requirement of long term habitability. Assuming $\tau_{hab}$ > 2 Gyr (see section "Definitions") we obtain a lower limit $N_{atm}$ > 300 g/cm$^2$ for an Earth-like value of $pCO_2$ (green curves). In this framework, a 3-fold decrease of the Earth's atmospheric mass would have made impossible the emergence of complex life on our planet. For the case $pCO_2$=38000 ppmv (red curves) we obtain $N_{atm}$ > 70 g/cm$^2$. This example of high $CO_2$ indicates that a low-mass atmosphere could still guarantee a long term habitability, but with a surface dose of radiation well above 100 mSv/yr, i.e. non optimal for complex life as we know it. The case with $p$CO$_2$=10 ppmv (blue curves) indicates that an atmosphere with low $p$CO$_2$ could sustain long term habitability only with if the total columnar mass of the atmosphere is sufficiently high, $N_{atm} > 1500$ g/cm$^2$. In this case, the surface dose would be safely below 100 mSv/yr.

The habitability times calculated above should be regarded as a lower limit to the true times of habitability if the atmospheric properties evolve in a such a way to compensate the gradual rise of insolation (possibly as a product of the presence of life, see e.g. Watson and Lovelock 1983). For instance, it has been suggested that the atmospheric background pressure may decrease during the evolution of $S$, possibly due to biological N$_2$ fixation and organic matter burial (Li et al. 2009). In this case, the ensuing smaller pressure broadening of greenhouse gases may extend the inner edge. In the case of the Earth this possibility has been questioned by the archean pressure measurements by Som et al. (2012), more consistent with an almost constant surface pressure for Earth. More likely, an active tectonic can help to widen the HZ by adjustment of the CO$_2$ level via the carbonate-silicate cycle (e.g. Kasting et al. 1993; Selsis et al. 2007; Kopparapu et al. 2013). The CO$_2$ abundance is expected to fade with increasing $S$ due to the rise of surface weathering with increasing temperature. The decrease of CO$_2$ can delay the onset of the greenhouse instability at the inner edge (Kasting 1988; Goldblatt & Watson 2012). However, $p$CO$_2$ should not fade below the above mentioned minimum thresholds for photosynthesis, i.e. $p$CO$_2$ ~ 10 ppmv for plants with C4 photosynthetic systems.

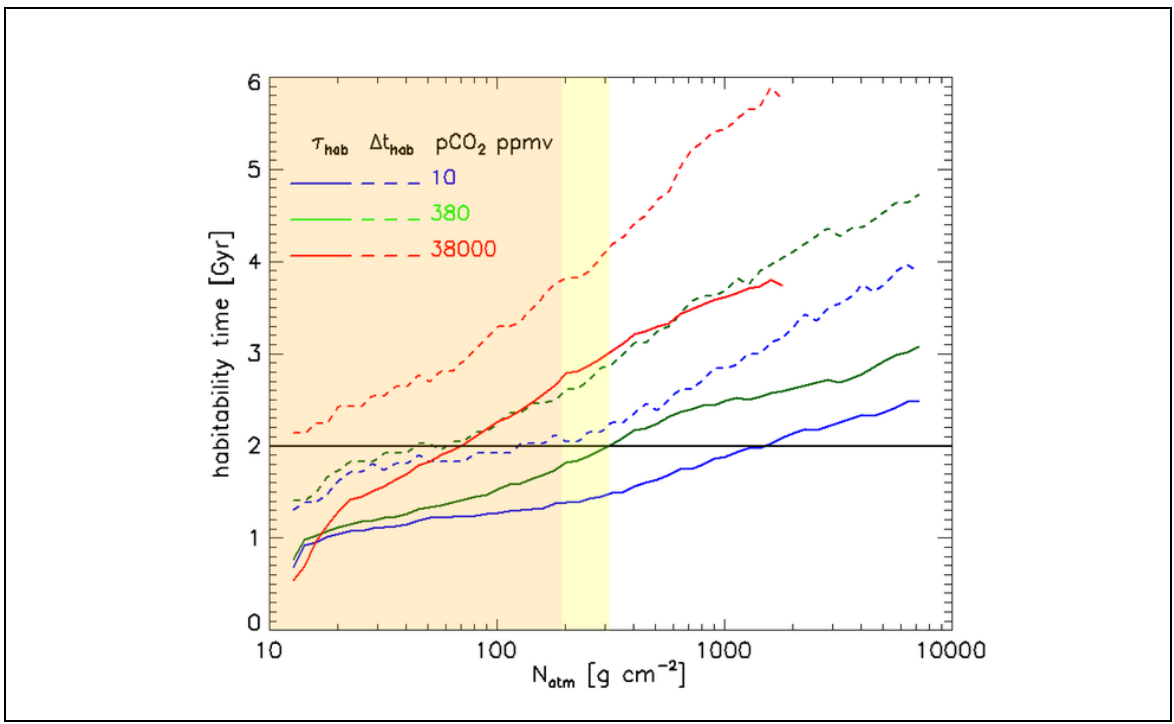

Figure 6 – Time span of habitability calculated with the "complex life" index h$_{050}$ for an Earth-like planet with different values of atmospheric columnar mass, N$_{atm}$, and CO$_2$ partial pressure, pCO$_2$. Times are calculated using the main sequence evolution of a sun-like star (Bressan et al. 2012), assuming that the planet lies at 1 AU from the star. Dashed curves: maximum time of habitability, $\Delta t_{hab}$. Solid curves: effective time of habitability, $\tau_{hab}$. Blue, green and red curves: pCO$_2$=10 ppmv, 380 ppmv, and 38000 ppmv, respectively. Horizontal line: representative time required for the emergence of complex life. The red curves are truncated where the insolation of the outer edge becomes lower than the minimum insolation at the initial main sequence (Fig. 3). See text.

If the carbonate-silicate cycle is at work, the gradual decrease of $p$CO$_2$ in the course of the planet evolution would yield a milder evolution of the latitude profiles of surface temperature and habitability. To illustrate this effect, in the bottom panels of Fig. 4 we show how the mean orbital profiles $\underline{T}(\varphi)$ and $\underline{H}_{050}(\varphi)$ would evolve when a rise of insolation ($S_{eff} = 0.77, 0.90, 1.00, 1.22$) is coupled to a decrease of $p$CO$_2$ ($f_{GH} = 100, 10, 1, 0.1$, respectively). One can see that the mean orbital temperature and habitability show less dramatic differences with respect to the case of constant $p$CO$_2$ ($f_{GH}=1$) shown in the upper panels of the same figure.

*Evolutionary tracks of complex life habitability*

By adopting a proper stellar evolutionary track we can explore the surface temperature and habitability as a function of planet age. Examples of application of this technique are presented in Fig. 7, where we show the evolution of $h_{050}$ and $T_m$ versus $S$ and $t^*$ for several representative climate models. Each of the 5 curves shown in figure corresponds to a constant value of $N_{atm}$ between 30 and 8000 g/cm$^2$, as specified in the legend. An Earth-like atmospheric composition, with a reference value $p$CO$_2$=380 ppmv, was adopted in all cases. The results show that the initial and final times of habitability are a strong function of $N_{atm}$. For planets with massive atmospheres the habitability starts and finishes relatively early, even though the time span of habitability (Fig. 6) is relatively large. For planets with low-mass atmospheres the habitability starts at much later stages and lasts for a relatively short time. It is not clear whether such a late start would be compatible with the emergence of life. For instance, in the case of the Earth, life emerged during the first Gyr after the formation of the Solar System and the conditions for the emergence of life may have faded at a later stage. For an Earth-like planet with present age 4.58 Gyr but different values of atmospheric mass, the epoch of habitability would have already passed for $N_{atm}>$ 4000 g/cm$^2$, while the right conditions would not have yet been met for $N_{atm} < 200$ g/cm$^2$, but these low values would imply a harsh radiation environment. Of course, as discussed for the time interval of habitability, also the actual initial and final times of habitability will depend on possible changes of the chemical composition and total mass of the atmosphere. For instance, by adopting present-day values of $N_{atm}$ and $p$CO$_2$, Fig. 7 indicates that an Earth twin would not be habitable for a long initial period, while we know that the primordial Earth was habitable since the archean. This is the well-known "faint young sun paradox" that can be solved assuming a high initial value of $p$CO$_2$ (see, e.g., Kasting 2003). An example is shown in Fig. 8.

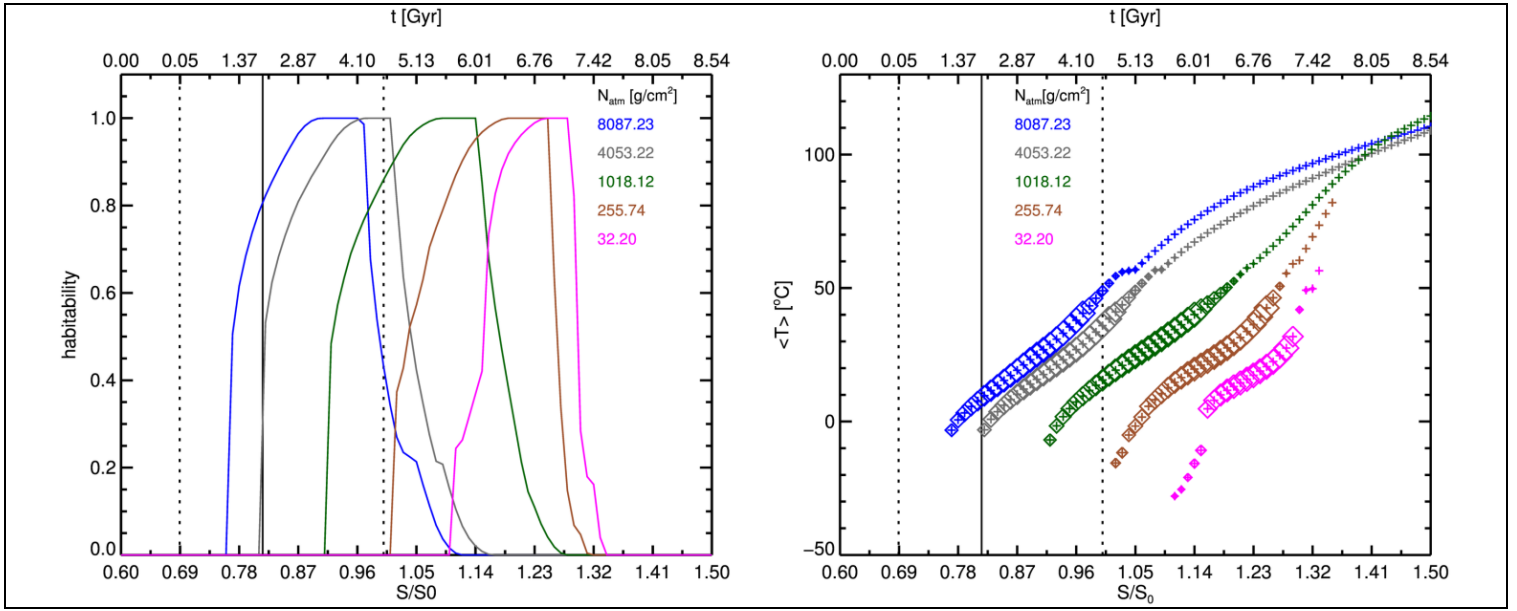

Figure 7 - Evolution of the mean global and orbital habitability $h_{050}$ (left panel) and temperature $T_m$ (right panel) as a function of planet insolation (bottom horizontal axis) and stellar age (top horizontal axis) for an Earth-like planet orbiting a sun-like star at a=1 AU. The stellar evolutionary track is taken from Bressan et al. (2012). The curves correspond to the 5 values of atmospheric columnar mass, $N_{atm}$, indicated in the panels. The dotted vertical lines indicate the age of arrival on the main sequence time, t=0.048 Gyr (the track includes the pre-MS evolution), and the current age, $t_0$=4.58 Gyr, respectively, the solid vertical line is the reference minimum $\tau_{hab}$ of 2 Gyr. The size of the squares on the right panel are proportional to the value of the habitability index $h_{050}$.

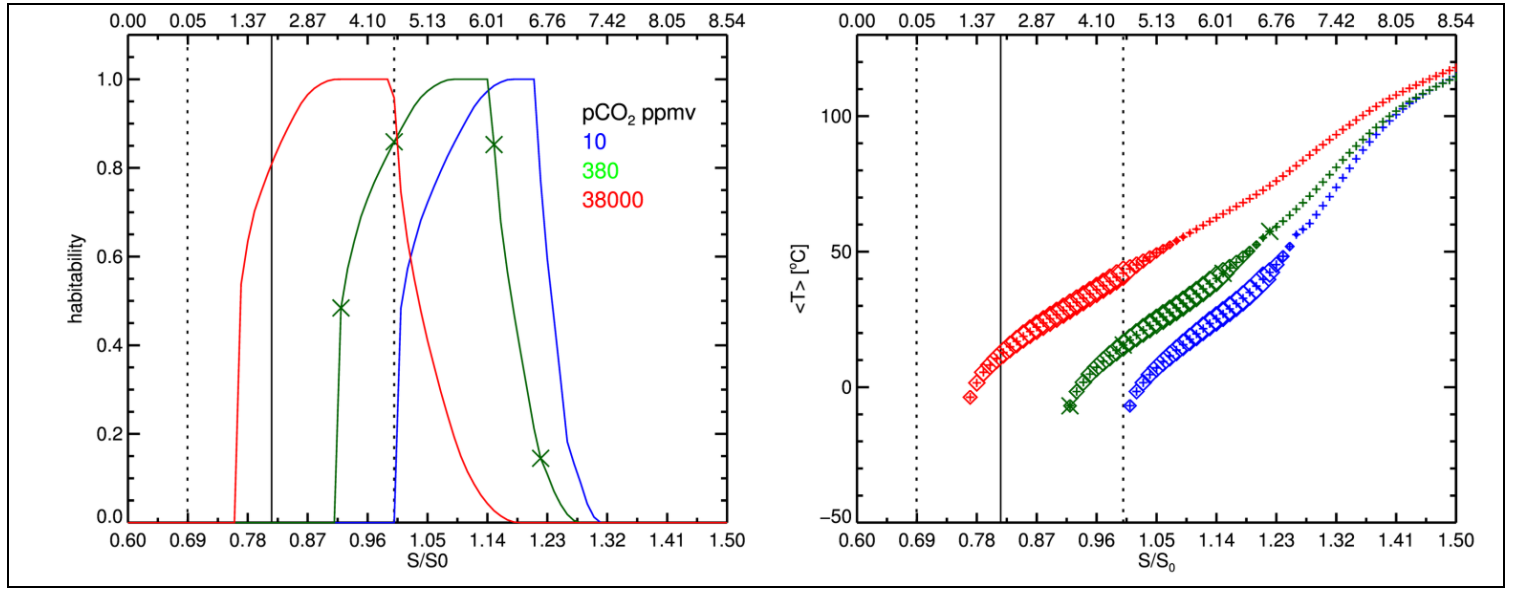

Figure 8 - Same as Fig. 7, but for fixed p=1 bar, and $pCO_2$=10 (blue), 380 (green), 380x100 (red) ppmv. The crosses on the green line correspond to the 4 models in the top panels of Fig. 4.

## Life evolution beyond the edges of the habitable zone

Different types of life may in principle exist outside the edges of habitability for complex life, depending on the previous evolution history. The situation is quite different at high insolation or at low insolation. We briefly discuss both cases.

Outside the outer edge, survival states of life with suspended metabolism may in principle exist and even be preserved for long periods of time. Taking terrestrial life as an example, survival states could be unicellular, such as spores, or even multicellular, as in the case of tardigrades. Whether these survival states do indeed exist or not depends on the previous history of the planet habitability. As we said, the planet is expected to displace through the AMHZ at roughly constant $N_{atm}$ and increasing $S$ during the main sequence evolution of its central star. In presence of a suitable level of geophysical activity, the existence of the carbonate-silicate cycle may gradually diminish $p(CO_2)$, shifting the edges of habitability towards higher values of insolation and keeping the planet inside the HZ (Walker et al. 1981). Whether the geochemical cycle is present or not, it seems unlikely that, in general, the rise of stellar luminosity would shift a planet from a location inside the HZ, where life might have developed, to a colder location outside the outer edge of habitability. For this type of evolution to occur, a major change of the atmospheric properties or orbital parameters of the planet must take place. For instance, the planet should undergo a dramatic reduction of $N_{atm}$ and/or $p(CO_2)$. Mars is a likely example of an evolution of this type, since the initial $CO_2$-rich atmosphere, invocated to explain many observed features, must have been lost in its early history, possibly as a result of the interaction with the solar wind (Jakosky et al. 2015). Only in situations of this type we may expect to find survival states of life outside the outer edge. Just outside the outer edge, cryophiles with metabolic activity down to $T \sim -20$ °C might also exist.

The situation is quite different for a planet crossing the inner edge. This is the natural ending for a planet with roughly constant $N_{atm}$ and increasing $S$, even in presence of a negative feedback of $p(CO_2)$ activated by the carbonate-silicate cycle. If the planet had a biosphere, thermophilic life may still survive for a relatively short period of time between the onset of the runaway greenhouse instability and the final loss of water from the planet. After water loss, some forms of (hyper)-thermophile life can in principle be maintained in semi-dry conditions, but only up to $T \sim 120$°C. Above this threshold the planet surface would be sterilized. We refer to O'Malley-James et al. (2013, 2014) for a detailed discussion of the last stages of life before the onset of the runaway greenhouse instability. In a land-dominated planet the high thermal limit could be approached without triggering the runaway greenhouse instability (Abe et al. 2011; Zsom et al. 2013). In this case the persistence of life inside the classic inner edge would be dictated by the resilience of (hyper)-thermophile life, rather than by the time scale of climate instabilities.

## Implications for the generation of atmospheric biosignatures

The generation of atmospheric biosignatures requires the widespread presence of life. The example of the Earth shows that unicellular life is a predominant constituent of the biosphere and has a major chemical impact on the atmospheric composition. Indeed, the predominance of unicellular life is expected to be a common feature of any life-bearing environment, because unicellular life has a greater flexibility of adaptation and lower energy requirements than multicellular life. In spite of this, multicellular life can have a significant impact on the atmosphere composition, as shown by the example of the plants on Earth. As we have discussed before, the thermal limits for plants are in line with those of other multicellular poikilotherms, i.e. $0$°C $\leq T_L \leq 50$°C. We now discuss the thermal limits of unicellular organisms in order to understand if the same limits are also relevant for the generation of atmospheric biosignatures by unicellular life.

The thermal limits of unicellular life are very broad, covering the interval $-20$°C $\leq T_L \leq 122$°C (Clarke 2014). However, at present time, extremophiles are only present in well defined environmental niches and therefore most of the impact on the atmosphere is provided by other, more widespread unicellular organisms. Methanogens can live in a broad spectrum of temperatures and are widespread in anaerobic conditions, yielding a signature of $CH_4$. However, the presence of complex life requires oxygen (e.g. Catling et al. 2005) and, as soon as oxygen rises, anaerobic life becomes less widespread, and the amount of methane in the atmosphere decreases due to $CH_4$ reacting with OH produced by $O_3$ photolysis and water (Kaltenegger et al. 2007, Rugheimer et al. 2013). At this stage cyanobacteria, which have both aerobic and anaerobic capabilities, become an essential contributor of atmospheric signatures through oxygenic photosynthesis. Cyanobacteria are extremely diffuse in a huge variety of terrestrial environments. The thermal limits $0^{\circ}C \leq T_L \leq 50^{\circ}C$ bracket most of the ambient temperatures of their habitats. Cyanobacteria in geothermal habitats represent a notable exception, since they can live up to 60 $^{\circ}C$ and even 70$^{\circ}C$ (Ward and Castenholz 2000). However, the number of cyanobacteria species in geothermal pools declines steeply above 50$^{\circ}C$ (see Fig. 3 in Clarke, 2014), suggesting that the photosynthetic contribution of thermophilic cyanobacteria is modest even in high temperature environments.

Summarizing the above experimental data, we conclude that the limits $0^{\circ}C \leq T_L \leq 50^{\circ}C$, that we adopt for index of complex life habitability $h_{050}$, are sufficiently broad for the generation of a chemical disequilibrium in the atmosphere by unicellular organisms and plants. Higher ambient temperatures could provide a significant contribution of atmospheric oxygen only in a planet where, at variance with the earth, thermophilic photosynthetic organisms dominate over mesophilic ones.

## Conclusions

The extreme technical difficulty in observing the atmospheres of rocky exoplanets, and the yet unknown origin of life on earth, make it hard to address the quest for life outside the Solar System. To focus this search, it is necessary to find ways of selecting the best candidate exoplanets with environmental (astronomical and planetary) conditions potentially capable of hosting life. With this aim in mind, we have introduced a temperature-based surface habitability index illustrative of planetary conditions suited for the potential emergence of complex life and generation of atmospheric biosignatures in exoplanets. The main points of our work are summarized in the following:

*Based on definitions of life, complex life and habitability sufficiently broad to be applied in exoplanet research, we showed that the thermal sensitivity of multicellular poikilotherms is the most appropriate to set temperature limits for the potential presence of complex life in exoplanets. The thermal limits for active life and reproduction of multicellular poikilotherms on earth lie in the range ~0-50°C. These limits are likely the result of universal, temperature-dependent mechanisms that affect all basic life processes, from the subcellular level up to the higher functions typical of complex life. The thermal limits become more stringent with increasing complexity of the organisms.

*We introduced the $h_{050}$ habitability index for complex life, defined as the orbital-averaged planetary surface fraction satisfying the 0-50°C temperature constraints. These thermal limits are sufficiently broad for the photosynthetic production of atmospheric oxygen, a key ingredient of complex life. Planets that are habitable according to the index $h_{050}$ can in principle be detected through observations of atmospheric biosignatures.

*With the help of the ESTM climate model, tailored for earth-like planets, we calculate $h_{050}$ as a function of planet insolation, $S$, and atmospheric columnar mass, $N_{atm}$, two of the physical quantities that most affect the planet surface temperature. By fixing the remaining planetary parameters, we build up the atmospheric mass habitable zone (AMHZ) for complex life in the $S$-$N_{atm}$ plane. For the sake of comparison with the liquid water criterion used in most studies of habitability, we also calculated the AMHZ with a pressure-dependent liquid water index, $h_{lw}$.

*The complex-life AMHZ, calculated with the $h_{050}$ index, shifts to lower values of insolation, as $N_{atm}$ increases. All the regions that are habitable according to the $h_{050}$ index have insolations below the runaway greenhouse (RG) limit and may thus benefit long term habitability. The inner edge of the $h_{050}$ AMHZ is not affected by the uncertainties inherent to the calculation of the RG instability that defines the inner edge of the classic habitable zone. Therefore, the difficulty of defining exact thermal thresholds for life is repaid by the advantage of obtaining results which are more robust from the point of view of climate calculations.

*The width of the complex-life AMHZ rises modestly with increasing $N_{atm}$, with a tendency to become stable (or even decrease) when $N_{atm} > 1000$-$2000$ g/cm$^{-2}$. Conversely, the liquid-water AMHZ, calculated with the $h_{lw}$ index, widens significantly with increasing $N_{atm}$, but with regions of high insolation that can be habitable only for a short period of time after the onset of the RG instability. Planets in such regions of high insolation are not suitable to host complex life and have a low chance of yielding oxygenic biosignatures, even if they are sufficiently dry to escape the RG instability (Abe et al. 2011, Zsom et al. 2013).

*The region of the AMHZ with low columnar mass, $N_{atm} < 300$ g/cm$^{-2}$, is only slightly dependent on $N_{atm}$. Below such threshold, a decrease of the atmospheric mass will not help the planet to become habitable. The properties of the AMHZ depend on the amount of $CO_2$, which we have fixed in trace amounts in each set of calculations. The AMHZ tends to widen and to shift to low insolation with increasing $pCO_2$. For some particular combinations of $N_{atm}$ and $pCO_2$ the width of the AMHZ tends to decrease with increasing $N_{atm}$. Further calculations are required to cast light on the width of the habitable zone in $CO_2$-rich atmospheres, not treated in the present work.

*The type of life potentially present inside the AMHZ changes according to its position on the $S$-$N_{atm}$ plane, which affects the mean global temperature, as well as the latitudinal gradients and the seasonal excursions of surface temperature. Large meridional gradients may favor the emergence of a spectrum of life forms as far as the thermal response in concerned. In the case of uniform $T$ profiles, the distribution of $|T_{opt} - T|$ would play a key role in determining the diversity of life. Planets with low (high) $N_{atm}$ are predicted to be subject to large (small) seasonal excursions, affecting the potential geographical distribution of life. Zones with strong excursions may favor the emergence of species with mechanisms of thermal adaptation and/or capability of meridional migration.

*The atmospheric columnar mass $N_{atm}$ determines the shielding of the planet atmosphere from external ionizing radiation, affecting the radiotolerance characteristics of life potentially present inside the AMHZ. We used calculations by Atri et al. (2013) to estimate the amount of biological-damaging radiation induced by Galactic cosmic rays reaching the surface. Planets with $N_{atm} < 200$-$300$ g cm$^{-2}$ would be subject to a radiation dose $>100$ mSv/yr, which is considered to be hazardous for humans in the long term.

*By coupling the impact of $N_{atm}$ on the surface temperature and radiation dose, we speculate that the large temperature excursions and high radiation dose expected at $N_{atm} < 200$-$300$ g cm$^{-2}$ may yield a fast rate of darwinian evolution triggered by the need of coping with such environmental challenges. The small seasonal excursions and low radiation doses expected when $N_{atm}$ is a few times the earth's value (1033 g/cm$^2$) may induce a lower evolutionary pressure.

*By using the $h_{050}$ index in combination with a stellar evolutionary track it is possible to calculate the habitability evolution for complex life. Since the habitability evolves within the HZ, we have defined a habitability weighted time scale, $\tau_{hab}$, in alternative to the total time span. The habitability timescales depend on $N_{atm}$ and the $CO_2$ abundance. Increasing $N_{atm}$ and $pCO_2$ shifts the epoch of habitability to earlier phases of stellar evolution. At $t^*=4.58$ Gyr an earth like planet with $N_{atm} \sim 4000$ g cm$^{-2}$ would be on the verge of exiting its HZ, while for $N_{atm} < 250$ g cm$^{-2}$ it would not had entered yet. The time span of habitability tends to rise as $N_{atm}$ increases. However, such calculations do not take into account the possibility that life itself could trigger atmospheric composition changes that can counterbalance the changes in stellar luminosity.

*It is possible to constrain the atmospheric columnar mass of the planet by requiring a minimum of $\sim 2$ Gyr of habitability conditions before the emergence of complex life. We find that this requirement implies $N_{atm} > 300$ g cm$^{-2}$ for a planet with an earth-like composition, and $N_{atm} > 70$ g cm$^{-2}$ for a 100 fold increase of $CO_2$ (however the surface would be subject to a very intense radiation dose below 200 g cm$^{-2}$). For 10 ppmv of $CO_2$, the right conditions would be met only for $N_{atm} > 1500$ g cm$^{-2}$.

Progress towards a full understanding of the effects of temperature across levels of biological organization will clarify the potential of terrestrial life as a reference in the quest for complex life in the universe. The methodology introduced in the present work will be useful to select the best candidate targets for future searches for life in exoplanets. Exploring the habitability for complex life for a wide range of planetary and stellar parameters will help to provide quantitative criteria for future studies of the Galactic habitable zone.

**Acknowledgements**

We thank Daniela Billi and Marco Moracci for helpful comments on the thermal limits of multicellular life and Alessandro Bressan for providing the stellar evolutionary track used in our calculations.

**Appendix**

**Atmospheric columnar mass and surface dose of cosmic rays radiation**

The mass of an atmospheric column of unit area, $N_{atm}=p/g$, impacts on the climate through its effects on the vertical and horizontal energy transport (see Section "The atmospheric mass habitable zone"). Besides these effects, $N_{atm}$ influences the habitability through the shielding effect of the atmosphere on ionizing radiation from space. For a given flux of ionizing radiation hitting the outer layers of the atmosphere, the surface dose of radiation will decrease with increasing $N_{atm}$. Since $N_{atm}$ is accounted for in the climate calculations of habitability, it is worth investigating its effects on the surface radiation dose. We focus our discussion on ionizing radiation induced by cosmic rays.

*Radiation dose induced by cosmic rays*

Cosmic rays are high-energy charged particles of astrophysical origin that can affect planetary habitability (Ferrari and Szuszkiewicz 2009; Atri and Melott 2014). The planet surface is hit by secondary particles resulting from the interaction of cosmic rays with the upper layers of the atmosphere. The resulting surface radiation dose may cause biological damage (e.g. Magill and Galy 2005). The central star plays a double role as far as cosmic rays are concerned. On one hand, the stellar winds blow an astrosphere that can shield the planet from Galactic cosmic rays (GCRs). On the other hand, the star itself can produce stellar flares that generate cosmic rays. The extent of the astrosphere and thus its protecting effect from GCRs can decrease during the motion of the planetary system across the galaxy, e.g. when passing through dense interstellar clouds or if hit by the shockwave from a SN explosion. Such descreening may affect planets distant from the central star, such as HZ planets of solar type stars (e.g. Scalo 2009). In fact, this phenomenon may have occurred during the geological history of earth (e.g. Fields et al. 2008; Zank & Frisch 1999). The descreening by GCRs strongly decreases for close-in HZ planets around M-type stars. In this case, however, the main risk is the central star, due to the frequent and intense flares typical of M-type stars, that may last for one billion years after star formation or even more (e.g. Griessmeier et al. 2005). A solar type star is expected to undergo a shorter active period (Ribas et al. 2005) and strong solar events occur seldom as it ages. For these reasons, planets in the HZ of solar-type stars are mainly subject to GCRs for most of the time spent in the main sequence.

The planet surface can be protected by cosmic rays of whatever origin through the shielding effect of the planet's magnetosphere and atmosphere, if present. A fast rotation rate is one condition for the existence of the dynamo effect that generates the planetary magnetic field. Planets in the HZ of M-type stars may lack magnetosphere since their rotation is braked by tidal interactions with the nearby central star. Planets in the HZ of solar-type stars are substantially less braked and may benefit from the protection of their magnetosphere. Here we consider habitable planets around solar-type stars, parametrizing the strength of their magnetic field in terms of the earth's value. We assume that GCRs are the dominant source of ionizing radiation and we consider the shielding effect of the planet atmosphere (Griessmeier et al. 2005; Atri et al. 2013). The cosmic rays that strike the upper atmospheric layers produce penetrating secondary particles that can hit the planet surface, such as muons, electrons, neutrons and photons. For a given flux of GCRs and a given planetary magnetic moment, the surface flux of secondary particles can be calculated as a function of atmospheric columnar mass, $N_{atm}=p/g$. The flux of secondary particles can be converted into energy deposition rate (J m$^{-2}$ s$^{-2}$) and thus into radiation dose (Sv/year), using standard factors of biological damage (Magill and Galy 2005). Here we adopt the results of the calculations provided by Atri et al. (2013) for planets with different values of $N_{atm}$ and magnetic moment.

*Radiation dose limits of complex terrestrial life*

Given the dependence of the surface dose on $N_{atm}$ we may, in principle, establish limits of habitability by assigning a threshold of radiation dose for life. However, assigning thresholds of radiation dose is even harder than assigning thermal thresholds. Within certain limits, terrestrial life has the capability of repairing the biological damages produced by ionizing radiation (Magill and Galy 2005). The degree of radio tolerance varies significantly among organisms, being larger in prokaryotes. The most famous case is the bacterium *Deinococcus Radiodurans*, which tolerates doses in excess of ~$10^4$ Sv (Stan-Lotter 2007). As in the case of the thermal limits, complex life has narrower limits of radio tolerance with respect to unicellular life. However, it is quite hard to assign a specific limit of radiation dose above which complex organisms cannot complete their life cycle. In part this is due to the paucity of experimental data for complex life, which are often obtained from studies of the impact of nuclear accidents. Assigning thresholds is also difficult because the biological damage depends not only on the dose, but also on its persistence over time. Typical values of lethal dose for humans lie in the range between 1 Sv and 10 Sv over short periods of time (Magill and Galy 2005). However, if the irradiation is persistent over long periods of time, a significant biological damage can result also at low doses. This regime of long-term irradiation is the one we are interested in, because the surface flux of secondary particles is persistent owing to the long term variation of the flux of GCRs, the planetary magnetic field, and the atmospheric columnar mass. In this regime of persistent, low dose it is hard to decide a threshold of tolerance because the biological damage becomes apparent only after integration over a significant period of time. As a result, it is difficult to gather a statistical sample to study the survival probability. Here we adopt a reference value of 100 mSv/yr, which is the international limit for nuclear workers. For comparison, the maximum permissible dose for the general public is of 1 mSv/yr (Magill and Galy 2005). The radiation dose rate on earth from all natural sources is about 2.4 mSv/yr, of which about 0.39 contributed by cosmic rays (Atri & Melott 2014). With 100 mSv/yr, a critical value in the order of 1 Sv would be reached in 10 yr, a time scale comparable to the life time of terrestrial metazoans.

It should be clear that the value of 100 mSv/yr is only an illustrative value for the long term habitability of complex life of terrestrial type.

Address correspondence to:
Laura Silva
INAF-OAT
Via G.B. Tiepolo 11
34143 Trieste
Italy
E-mail: silva@oats.inaf.it